\documentclass[twocolumn,aps,prb,showpacs,raggedbottom,nobalancelastpage,amssymb,superscriptaddress]{revtex4-1}

\usepackage{amsmath}
\usepackage{amssymb}
\usepackage{color}
\usepackage{graphicx}

\newcommand{\bra}[1]{\left<{#1}\right|}
\newcommand{\ket}[1]{\left|{#1}\right>}
\newcommand{\braket}[2]{\left<\left.{#1}\right|{#2}\right>}

\begin{document}
\title{Measuring cotunneling in its wake}

\author{Oded Zilberberg}
\affiliation{Institute for Theoretical Physics, ETH Zurich, 8093 Z{\"u}rich, Switzerland}
\affiliation{Department of Condensed Matter Physics, Weizmann Institute of Science, Rehovot 76100 Israel.} %
\author{Assaf Carmi}
\affiliation{Department of Condensed Matter Physics, Weizmann Institute of Science, Rehovot 76100 Israel.} %
\author{Alessandro Romito}
\affiliation{\mbox{Dahlem Center for Complex Quantum Systems and Fachbereich Physik, Freie Universit\"at Berlin, 14195 Berlin, Germany}}

\begin{abstract}
We introduce a rate formalism to treat classically forbidden electron transport through a quantum dot (cotunneling) in the presence of a coupled measurement device. We demonstrate this formalism for a toy model case of
cotunneling through a single-level dot while being coupled to a strongly pinched-off quantum point contact (QPC).
We find that the detector generates three types of back-action: the measurement collapses the coherent transport
through the virtual state, but at the same time allows for QPC-assisted incoherent transport, and widens the dot
level. Last, we obtain the measured cotunneling time from the cross correlation between dot and QPC currents.
\end{abstract}

\maketitle
\section{Introduction}

Quantum measurement is a probabilistic process where the
detector’s outcome is correlated with the system being in
a certain state. In turn, the detector’s back-action onto the
system affects it according to the specific outcome of the
measurement~\cite{Helstrom}.
 In a projective measurement this is described
by the wave-function collapse, in which the outcome of the
measurement is an eigenvalue of the measured observable,
and the system after the measurement is projected onto the
eigenstate corresponding to the obtained eigenvalue~\cite{von-neumann}.

The implications of quantum measurement become particularly interesting when applied to yet another key feature
of quantum mechanics--—to classically forbidden processes.
A striking example of such a forbidden process is that of
tunneling under a potential barrier, where a particle appears
on the other side of a barrier that it classically could not
surmount. Whereas this appearance indicates that a tunneling
event occurred, a direct observation of the particle during
its virtual passage under the barrier is required as additional
verification of this mechanism.

Indeed, such a proposition stands in direct conflict with
quantum measurement: if the particle is measured to be under
the barrier, it would collapse midway and not tunnel to the
other side. Additionally, in order for the particle to collapse in
the position of the barrier, it must obtain energy and appear
above the barrier. As a result, one may conclude that such a
measurement would correspond to effectively increasing the
potential barrier, and blocking tunneling altogether. Still there
is a constant interest in accessing the properties of a particle
tunneling under a barrier (e.g., its traversal time), with a variety
of approaches~\cite{Condon1931,Wigner1955,Buttiker1982,Sokolovski1987,Steinberg1995}.

Beyond the scheme of projective measurement, a better
suited and more realistic approach in detecting the virtual state
under the barrier is that of a modification of the systems’s
state in a continuous process accompanied by a gradual
acquisition of information by the detector~\cite{Clerk2010}. 
In particular,
in a weak measurement regime, as opposed to a strong
projective measurement, the detector’s outcome corresponds
to the measured state of the system, but the back-action does
not disturb it much. This allows for nontrivial effects in
conditional (postselected) measurements, e.g., the appearance
of weak values~\cite{Aharonov:1988a}, and utilization of the measurement outcome
in quantum feedback circuits \cite{Korotkov:2001a,Korotkov:2001b}. 
Such effects have been successfully employed in practical problems including
precision measurements~\cite{Hosten:2008,Dixon:2009,Starling:2009,Brunner:2010,Starling:2010b,Zilberberg:2011}, quantum state discrimination~\cite
{Zilberberg2013} or quantum state stabilization~\cite{Korotkov:2001b}. Moreover,
weak measurements have also been successfully employed
in the study of coherent quantum transport under a potential
barrier~\cite{Buttiker1983,Steinberg1995}, as well as transport through many-body virtual states~\cite{Romito2014}.

The detection of a tunneling process via weak measurements can be directly explored in electronic solid-state devices.
The typical system under consideration is that of transport
between two leads across a quantum dot. By tuning the
capacitance of the dot, one can position its eigenenergies
relative to the chemical potential of the leads, such that an
addition of charge onto the dot is unfavored by Coulomb
interactions. As a result, transport through the dot is classically
blocked~\cite{Aleiner2002}. In this regime, the transport through the dot
happens via cotunneling processes with a virtual (classically
forbidden) occupation of the dot~\cite{Glazman2005}. 

Charge detection in quantum dots using a quantum point
contact (QPC) that is capacitively coupled to the dot is
experimentally well established~\cite{Field1993,Elzerman2003,DiCarlo2004,Harbusch2010,Gasparinetti2012,Granger2012,clemens2012,clemens2013,zumbuhl2014}. The QPCs can be tuned to be close to the quantum limit, making them
sensitive to the dot’s charge fluctuations. Theoretically QPC
detectors are well understood, and a proper formalism has
been developed to describe the many-electron macroscopic
classical signal therein in response to local charges~\cite{gurvitz,Korotkov:2001b}.
Yet these descriptions are developed for classical transport
through the dot~\cite{gurvitz}, or for coherent charge oscillations in
isolated systems, e.g., double quantum dots~\cite{gurvitz,Korotkov:2001b,Romito:2008}.
In a recent work, measurement of cotunneling was addressed in the
regime of weak coupling between a dot and a QPC~\cite{Romito2014}. Using
a weak value approach extended to deal with the interacting dot
and a large bandwidth of the QPC, the short cotunneling time
was resolved. This led to a vanishingly small cotunneling time
in the regime of a diffusive 2D dot with sufficiently closed
contacts.

In this work, we extend the standard model of QPC transport~\cite{gurvitz,Korotkov:2001b} to treat the partial measurement of virtual
occupation of the dot in the cotunneling regime. In our
scheme, we treat the interaction between quantum dot (QD)
and QPC exactly by utilizing a rate equation formalism, which is perturbative in the tunneling and exact in the regime
of an almost pinched-off QPC. We illuminate and study
three mechanisms of back-action that a detector induces
onto coherent transport, namely (i) increased phase space for
QPC-assisted transport, e.g., inelastic processes in which an
electron enters the dot with energy $\epsilon$ and leaves the dot with energy $\epsilon' \neq \epsilon$ while another electron crosses the QPC such
that the total energy is conserved, (ii) reduced elastic transport
(decoherence), and (iii) widening of the dot energy levels.
Thus, we can address some of the questions introduced above:
we find that due to (i) and contrary to the prediction above,
the measurement amplifies the current through the dot and
does not block it. Nonetheless, mechanism (ii) implies that
transport would be blocked if the dot is incorporated inside an
interferometer.

Alongside these main results, we find that the detector
signal shows nonmonotonous behavior as a function of the dot
variables, which we attribute to sensitivity to the directionality
of transport through the dot. Additionally, we determine the
correlation between the QPC transport and the successful
cotunneling passage through the quantum dot. Interestingly,
despite the fact that the regime considered here is different
from that of previous approaches which established such a
relationship~\cite{Steinberg1995,Romito2014}, we can use the correlated-currents signal
to extract the time of the cotunneling process. The obtained
time is compared with the simple Heisenberg’s uncertainty
expectation of $\tau_{\rm{cot}} \sim \hbar/\Delta \mathcal{E}$, which is set by the inverse of the
``energy-debt'' during the virtual transport.

Our results are relevant for contemporary transport experiments~\cite{Harbusch2010,clemens2012}, where different mechanisms of QPC backaction are discussed and an estimation of the cotunneling time
is obtained from the linewidth of the differential conductance
through the dot. Note that, differently from other approaches
dealing with the many-body physics in the cotunneling system
and detector~\cite{Romito2014}, our formalism addresses the regime where
single cotunneling events are correlated with single electron
signals through the QPC.

The paper is structured as follows: in Sec.~\ref{model}, we write down the model of a dot measured by a QPC. In Sec.~\ref{method}, we derive the essential tools and methods for describing the measurement of cotunneling in its wake. Section \ref{result}, details the resultant interplay between dot and QPC, and its implications on measurable currents on correlations, as well as the ability of measuring the cotunneling time. In Sec.~\ref{conclude}, we conclude and discuss possible future directions. We provide a comprehensive Appendix \ref{rigorous}, which details the microscopic calculation required to establish the formalism used in Sec.~\ref{method}.

\section{Model}
\label{model}
Our setup is divided into two components: the system, and the detector that measures it,
\begin{align}
	H=H_{\text{sys}}+H_{\text{det}}+H_{\text{int}}\, ,
\end{align}
where $H_{\text{sys}}$ describes the system, $H_{\text{det}}$ describes the detector, and $H_{\text{int}}$ describes the interaction between them. For simplicity, we consider a spinless problem. The results can be directly extended in some regimes to the spinful case.
 \begin{figure}
 \centering
 \includegraphics[width=0.9\columnwidth]{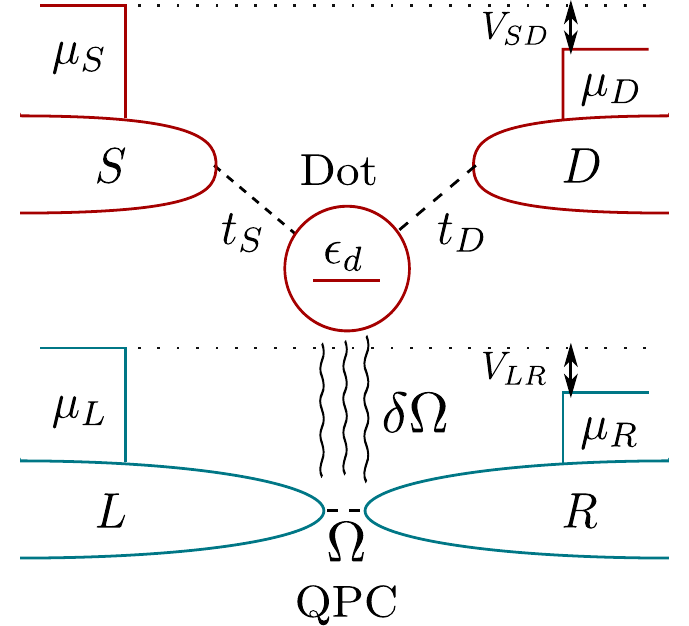}
 \caption[]{\label{Fig:1}
 A sketch of the setup: a single-level dot is tunnel-coupled to two leads, $S$ and $D$ with amplitudes $t_S$ and $t_D$, respectively. The respective dot-leads chemical potentials are $\mu_S$ and $\mu_D$, corresponding to a voltage bias $eV_{SD}=\mu_S-\mu_D$. A quantum point contact (QPC) is capacitively coupled to the dot, i.e., transport between the leads $L$ and $R$ is governed by a tunneling amplitude $\Omega$ when the dot is empty, and $\Omega-\delta \Omega$ when the dot is occupied. The respective QPC-leads chemical potentials are $\mu_L$ and $\mu_R$, corresponding to a voltage bias $eV_{LR}=\mu_L-\mu_R$.
 }
 \end{figure}
The system consists of a single level quantum dot, that is
tunnel-coupled to two electronic leads, the source $S$, and the
drain $D$ (see Fig. \ref{Fig:1}). 
By applying a voltage bias, $eV_{SD}=\mu_S-\mu_D$, between the source
and the drain chemical potentials ($\mu_S$, $\mu_D$), we can measure
the transport properties of the dot. We assume, henceforth, that $eV_{SD} \geq 0$.
The quantum dot is capacitively coupled to another lead, the gate
lead. 
Varying the gate-voltage, $V_g$, on the gate lead, controls the number
of electrons in the dot. 
The Hamiltonian that describes the system is
\begin{equation}
\label{eq: H}
H_{\text{sys}}=H_{SD}+H_{\text{dot}}+H_T\, ,
\end{equation}
where
\begin{align}
H_{SD}&=\sum_{k,\alpha=S,D}\epsilon_{k,\alpha} c_{k,\alpha}^\dagger c_{k,\alpha}\, ,\\
H_{\text{dot}}&=\epsilon_d d^\dagger d\, ,\\
\label{eq: Ht}
H_{T}&=\sum_{k,\alpha=S,D}t_\alpha c_{k,\alpha}^\dagger d + \text{H.c.}\, .
\end{align}
The operator $c_{k,\alpha}$ annihilates an electron with momentum $k$
and energy $\epsilon_{k,\alpha}$ in the lead $\alpha\in\{S,D\}$, $d$
annihilates an electron on the dot with energy $\epsilon_d$, which is
modulated by $V_g$. 
We have assumed that the tunneling coefficients, $t_\alpha$, between
the lead $\alpha$ and the dot are independent of the energy. 

As long as $\mu_S > \epsilon_d >\mu_D$, transport occurs via sequential
tunneling processes through the dot. 
When the dot level is outside the energy window provided by the leads, 
$\mu_D < \mu_S < \epsilon_d$ or $\mu_S > \mu_D > \epsilon_d$, and at low temperature, $T \ll \min\left\{\left|\mu_D-\epsilon_d\right|,\left|\epsilon_d-\mu_S\right|\right\}$, sequential tunneling is exponentially suppressed.
Nonetheless, a small current is still detected in the drain. 
This small current is carried by the so-called cotunneling processes,
in which electrons from the source virtually tunnel through the dot
into the drain~\cite{Averin1990,Glazman2005}.
By virtually we mean that the tunneling into the dot is classically
forbidden by energy conservation, but the overall cotunneling
process is energy conserving.  
In the present work we focus on this cotunneling regime.

We now turn to describe the detector. The detector is chosen to be a QPC because of its non-invasive nature \cite{Korotkov:2001a}. The QPC consists of two leads, left and right, that are tunnel-coupled to each other, see Fig.~\ref{Fig:1}. It is described by the following Hamiltonian
\begin{align}
 H_{\text{det}}=&\sum_l \mathcal{E}_l a_l^\dagger a_l+\sum_r \mathcal{E}_r a_r^\dagger a_r\nonumber\\
 &+\sum_{l,r}\Omega\left(a_l^\dagger a_r+\text{H.c.}\right)\, ,
\end{align}
where $a_{l}$ ($a_r$) annihilates an electron on the left (right) lead with momentum $l$ ($r$) and energy $\mathcal{E}_l$ ($\mathcal{E}_r$). The tunneling amplitude between the left and right leads, $\Omega$, is assumed to be energy independent. The QPC is out-of-equilibrium, namely, the chemical potentials of the left and right leads are different, $\mu_L=eV_{LR}/2$, $\mu_R=-eV_{LR}/2$.

The QPC is capacitively coupled to the quantum dot, i.e.~the tunneling amplitude, $\Omega$, is modulated by the charge on the dot
\begin{equation}
H_{\text{int}}=-\sum_{l,r}\delta\Omega \; d^\dagger d\left(a_l^\dagger a_r+\text{H.c.}\right)\, .
\end{equation}

The typical time scales for a tunneling event in the QPC to occur, are $\mathcal{D}^{-1}$ and $\tilde{\mathcal{D}}^{-1}$ for an empty and an occupied dot, respectively~\cite{gurvitz}. They are given by $\mathcal{D}=2\pi\Omega^2\rho_L\rho_R eV_{LR}/\hbar$ and $\tilde{\mathcal{D}}=2\pi\tilde{\Omega}^2\rho_L\rho_R eV_{LR}/\hbar$ where $\tilde{\Omega}\equiv\Omega-\delta\Omega$. For simplicity, henceforth, we assume the same density of states in the left and right leads, $\rho_L=\rho_R\equiv\rho$.


Note that a single-level quantum dot model corresponds to a quantum dot with appreciable level
spacing, larger than the other energy scales in the problem. Such a system can
be experimentally realized in semiconductors. Similarly, the QPC detector can be easily tuned to detect single electron tunneling in the sequential limit~\cite{Sukhorukov2007}, and to weak coupling in the cotunneling regime~\cite{clemens2012,clemens2013}. Furthermore, this model provides a valid effective description of multi-level quantum dots in the limit where the Thouless energy is the largest energy scale~\cite{Aleiner2002}.

\section{Methods}\label{method}
The currents through the dot and the QPC are carried by tunneling processes between the source and the drain via the dot, and from the left lead to the right lead in the QPC. In order to obtain these currents and their cross-correlations, we use a rate equation formalism~\cite{sakurai1985,Korotkov:1994,Koch:2006} and calculate the rates for these tunneling processes (see Appendix~\ref{rigorous}).
We assume that all the tunnel-couplings are weak, namely, $\rho_\alpha |t_\alpha|^2\ll eV_{SD}$ and $\rho|\Omega|^2\ll eV_{LR}$, and treat them perturbatively. Within this approach the interaction term, that makes the QPC transport dependent on the dot-occupancy, is treated exactly.

We focus on cotunneling rates, in which the occupation of the dot changes only virtually. Cotunneling can occur either between two different leads (e.g.~, source to drain) or back and forth between a lead and the dot (e.g.~, source to source). During these processes electrons can tunnel through the QPC. Therefore, the QPC current is sensitive to the virtual changes in the dot-occupancy. We denote by $W_{\alpha\alpha'}^{n}$ the cotunneling rate from lead $\alpha$ to lead $\alpha'$ (where $\alpha,\alpha'\in\{S,D\}$), during which $n$ electrons pass through the QPC.

\subsection{Toy model}
\begin{figure}
 \centering
 \includegraphics[width=0.5\columnwidth]{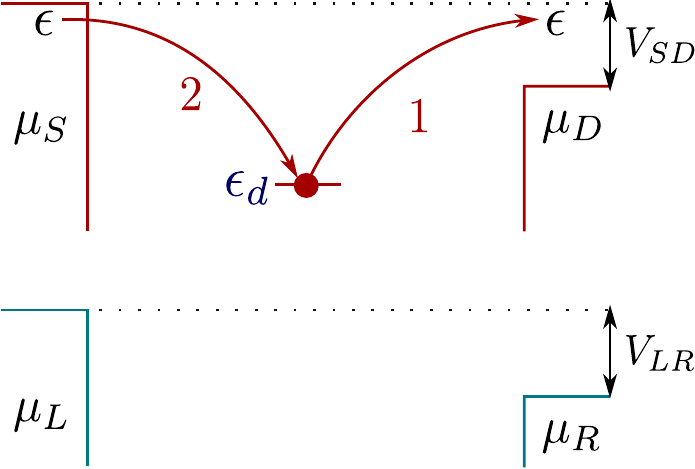}
 \caption[]{\label{Fig:2a}
 Sketch of the rate $W_{SD}^0$ of cotunneling through the dot accompanied by no tunneling through the QPC [cf.~Eq.~\eqref{eq: Wsd0}]. Similar to Fig.~\ref{Fig:1}, the upper part (brown) depicts the dot-system and the lower (blue) the QPC. }
 \end{figure}

\begin{figure}
 \centering
 \includegraphics[width=\columnwidth]{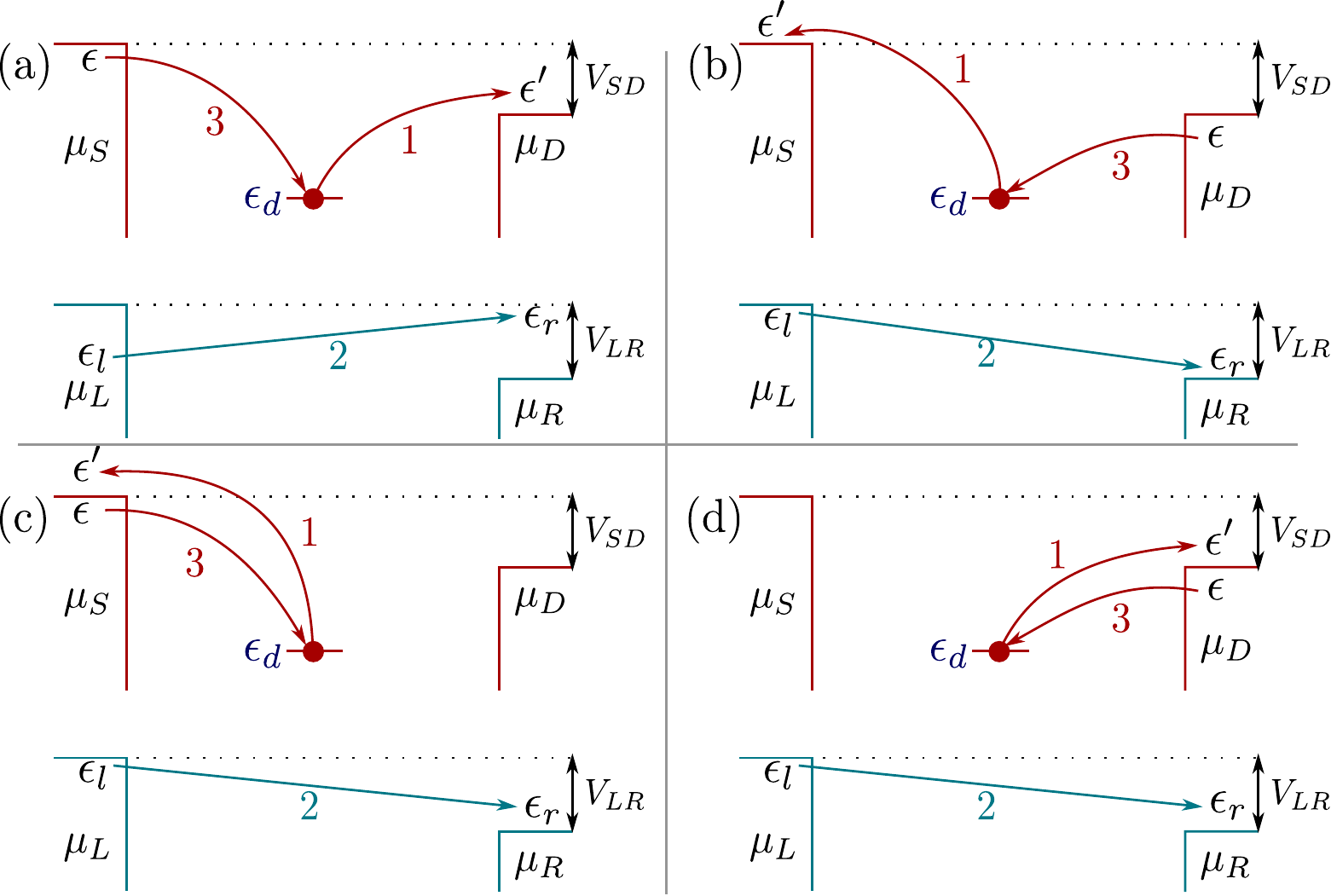}
 \caption[]{\label{Fig:2}
 Sketches of the relevant cotunneling rates in our toy model that involve an electron tunneling through the QPC [cf.~Eq.~\eqref{integralForm}]. Similar to Fig.~\ref{Fig:1}, the upper part (brown) depicts the dot-system and the lower (blue) the QPC. (a) The rate $W_{SD}^1$. (b) The rate $W_{DS}^1$. (c) The rate $W_{SS}^1$. (d) The rate $W_{DD}^1$. }
 \end{figure}

To study the interplay between the QPC and the dot we introduce a
simplified version of our model. In this simplified toy model, we wish to have, at most, one electron that tunnels through the
QPC during a cotunneling process through the dot, namely,
$W_{\alpha\alpha'}^n$ is negligible for $n>1$. 
Thus, only five rates remain relevant in the analysis of the currents
and cross-correlations: $W_{SD}^0$, $W_{SD}^1$, $W_{DS}^1$, $W_{SS}^1$, and
$W_{DD}^1$ [see Figs.~\ref{Fig:2a} and \ref{Fig:2}(a)-(d), respectively]. 

The rate $W_{SD}^0$ describes processes in which an electron
co-tunnels through the dot while no tunneling events through the QPC
take place. 
Therefore, this rate contributes only to the current through the
dot. The rates $W_{SD}^1$ and $W_{DS}^1$ describe processes in which, in addition to
cotunneling through the dot, an electron tunnels across the QPC. 
They contribute both to the current through the dot, and to the current
through the QPC. 
Hence, they generate cross-correlations between these currents. The rates
$W_{SS}^1$ and $W_{DD}^1$ are of processes that contribute to the
current through the QPC, but not to the current through the dot.  
Here, the electron tunnels back and forth between the dot and the same lead.

In order to work in the regime of our toy model, we take the following assumptions: First, we work in the zero temperature limit and assume that  no tunneling events occur in the QPC when the dot is occupied, i.e.,  we set
$\tilde{\Omega}=0$. As a result, transport through the QPC occurs only alongside
cotunneling events, i.e.~only during the time interval in which the
dot is virtually empty, $\tau_{\rm cot}$. 
We further take the limit where $\tau_{\rm cot}$ is much shorter than
the typical tunneling time in the QPC, i.e.~$\mathcal{D}\tau_{\rm
  cot}\ll 1$. 

The cotunneling time, $\tau_{\rm cot}$, is related by Heisenberg's uncertainty principle to the difference
between the energy of the initial (or final) state, and the energy of
the virtual interim state, e.g.~$\epsilon-\epsilon_d$ and
$\epsilon'-\epsilon_d$ in Fig.~\ref{Fig:2}. 
Of all cotunneling processes, the process with the longest
cotunneling time is the one with the lowest energy difference,
namely, for $\epsilon'=\mu_D$ (assuming $\mu_S>\mu_D$). 
Hence, the assumption above implies that
$\hbar \mathcal{D}\ll\mu_S-\epsilon_d,\;\mu_D-\epsilon_d~$.

Note that this chosen working point of the QPC puts the detector in a regime
different from the one typically considered in the context of weak
measurements.
In a typical weak measurement one assumes to work at finite
$\mathcal{D}$ and $\tilde{\mathcal{D}}= \mathcal{D}- \delta \mathcal{D}$.
The measurement process is then characterized by a measurement time,
$\tau_M \sim \mathcal{D}/\delta \mathcal{D}^2$, needed to resolve the
occupancy of the dot.
One can then work in a regime with $\tau_{\rm{cot}} \mathcal{D} \gg
1$ and $\tau_{\rm{cot}} \delta \mathcal{D} \ll
1$, such that the detector's bandwidth is larger than the cotunneling
time we are interested in ($\mathcal{D} \gg 1/\tau_\textrm{cot}$), but the
measurement during such a time remains weak ($\tau_{\rm{cot}} \ll
\tau_M$).

In our case $\delta \mathcal{D}=\mathcal{D}$, and as $\tau_{\rm{cot}} \mathcal{D} \ll
1$ we do not have a large bandwidth in the detector. 
Nonetheless, due to the very same choice, $\tilde{\mathcal{D}}=0$, corresponding to a blocked transport through the QPC, the current in the QPC is strongly correlated with a cotunneling through the dot, and one can expect that virtual transport through the dot could be resolved. In other words, the point $\tilde{\mathcal{D}}=0$ is peculiar since, as soon
as an electron tunnels through the QPC, one can infer with certainty
the happening of a cotunneling event through the quantum dot.
This is not a weak measurement in the standard sense, but resembles more a partial-collapse measurement~\cite{Katz2006,Korotkov:2007,Katz08,Zilberberg2012,Zilberberg2013}. Nonetheless, the rarity of such joint tunneling events generates limited back-action, and the detector can be used as a non-destructive  detector of the virtual occupation of the dot. In any case, this parameters' regime has the advantage of
simplifying the formalism, and rendering the processes involved in the
detection clearer~\cite{bandwidth}.

Experimentally, the small bandwidth of the QPC should not serve as a hurdle, as there are sufficiently many cotunneling events in the dot to obtain a measurable signal. We envision that tuning in to this measurement regime corresponds to starting from a conductance plateau in the QPC in the presence of a full dot that is decoupled from leads, i.e.~, no virtual charge fluctuation on the dot occurs. Increasing the coupling of the dot to its leads allows for cotunneling. If at the same time a QPC current is generated, it should correspond to the mechanism reported here.

\subsubsection{Derivation of the rates}
\label{derivationOfRates}
In Appendix \ref{tunnelRates}, we rigorously derive the rates
$W_{SD}^0$, $W_{SD}^1$, $W_{DS}^1$, $W_{SS}^1$, and $W_{DD}^1$. 
Here, we present a simpler, and intuitive way, to calculate these
rates.

We start by incorporating one of the effects of the coupling with the QPC by bestowing a finite width to the energy level of the dot, i.e.~the single level gains an effective width $\hbar(\mathcal{D}-\tilde{\mathcal{D}})/2$, which is proportional to the measurement strength. Note that in our toy model the measurement strength is, therefore, equivalent to $\hbar\mathcal{D}/2$. Hence, the cotunneling rate is simply the standard cotunneling rate through a single-level dot with energy $\epsilon_d$ and width $\hbar\mathcal{D}/2$. This is indeed the effect obtained by the microscopic calculation, which includes the dynamics of the QPC in Appendix \ref{tunnelRates}.

The total cotunneling rate $W_{SD}^0$, is the sum over the rates of all possible cotunneling processes (namely, integration over all incoming energies):
\begin{align}
&W_{SD}^0=\frac{2\pi}{\hbar}\int_{\mu_D}^{\mu_S}d\epsilon\; \rho_S \rho_D \left|\frac{t_St_D}{\epsilon-\epsilon_d-i\hbar\mathcal{D}/2}\right|^2\label{eq: Wsd0}\\
&=\frac{\Gamma_S \Gamma_D}{ \pi \mathcal{D}} \left[\tan^{-1}\left(\frac{2(\mu_S-\epsilon_d)}{\hbar\mathcal{D}}\right) -\tan^{-1} \left(\frac{2(\mu_D-\epsilon_d)}{\hbar\mathcal{D}}\right)\right]\;,\nonumber
\end{align}
where $\Gamma_\alpha\equiv2\pi\rho_\alpha|t_\alpha|^2/\hbar$. For later use, we also denote the total coupling $\Gamma=\Gamma_S+\Gamma_D$.

The cotunneling events through the dot, which are accompanied by a tunneling event through the QPC, consist of three stages (see Fig.~\ref{Fig:2}): First, the electron in the dot tunnels out into the source or the drain. Second, a tunneling event occurs in the QPC. Last, a new electron tunnels into the empty dot. Here too, we include an effective width, $\hbar\mathcal{D}/2$ to the energy of the dot. The total tunneling rate is given by
\begin{widetext}
\begin{align}
W_{\alpha\alpha'}^1&=\frac{2\pi}{\hbar}\int_{-\infty}^{\mu_{\alpha}}d\epsilon\int_{\mu_{\alpha'}}^{\infty}d\epsilon'\int_{-\infty}^{\mu_L}d\epsilon_l\int_{\mu_R}^{\infty}d\epsilon_r\; \rho_S \rho_D \rho_L \rho_R
\left|\frac{t_{\alpha}t_{\alpha'}\Omega}{(\epsilon'-\epsilon_d-i\hbar\mathcal{D}/2)(\epsilon_r+\epsilon'-\epsilon_l-\epsilon_d-i\hbar\mathcal{D}/2)}\right|^2\delta(\epsilon+\epsilon_l-\epsilon'-\epsilon_r)\nonumber\\
&=\frac{\hbar^2 \Gamma_{\alpha} \Gamma_{\alpha'} \mathcal{D}}{4 \pi^2  e V_{LR}}\int_{\mu_{\alpha'}-eV_{LR}}^{\mu_{\alpha}} d\epsilon\int_{\mu_{\alpha'}}^{\epsilon+eV_{LR}} d\epsilon'\frac{   (eV_{LR}+\epsilon -\epsilon ')}{\left[(\epsilon-\epsilon_d)^2+\frac{\hbar^2\mathcal{D}^2}{4}\right] \left[(\epsilon'-\epsilon_d)^2+\frac{\hbar^2\mathcal{D}^2}{4}\right]}\, .
	\label{integralForm}
\end{align}
\end{widetext}
The integrals in Eq.~\eqref{integralForm} are solvable. Out of space consideration, we choose to present these rates in integral form.
Here we notice the full extent of the interaction of the QPC with the dot; the QPC affects the cotunneling through the dot in two fashions: First, as previously noted, it gives a finite width to the energy level of the dot. Second, it allows new QPC-assisted cotunneling processes through the dot, where electrons are emitted from lead $\alpha$ \emph{inelastically} into lead $\alpha'$, provided that the energy difference is compensated by the QPC.

When the dot and the QPC are decoupled ($\Omega=0$) all the cotunneling processes are elastic, namely, an electron at energy $\epsilon$ in the lead $\alpha$, can be emitted into lead $\alpha'$ only at the same energy, $\epsilon$. In our case, the dot and the QPC are coupled ($\Omega\neq0$), and the two systems can exchange energy. Therefore, an electron with energy $\epsilon$ in lead $\alpha$ can end up in lead $\alpha'$ having a different energy $\epsilon'\neq\epsilon$. This is enabled by an electron with energy $\epsilon_l$ which is transmitted from the left lead of the QPC into the right lead with energy $\epsilon_r\neq \epsilon_l$ keeping the total energy conserved $\epsilon+\epsilon_l=\epsilon'+\epsilon_r$. Examples of inelastic cotunneling processes are schematically depicted in Fig.~\ref{Fig:2}. The total cotunneling rate $W_{\alpha\alpha'}^1$, is the sum over the rates of all possible cotunneling processes (namely, integration over all energies in the relevant leads of the dot and QPC systems).

At zero temperature, the maximal energy that a QPC-electron can lose
by tunneling is $\mu_L-\mu_R=eV_{LR}$ (assuming that
$\mu_L>\mu_R$). Hence, the maximal energy that a dot-electron can gain
is $eV_{LR}$, namely, $\epsilon-\epsilon'\leq eV_{LR}$ [see
Figs.~\ref{Fig:2}(b)-(d)]. Similarly, the maximal energy that the dot-electron
can lose is $eV_{SD}$ [see Fig.~\ref{Fig:2}(a)]. Increasing the applied bias voltages (both on
the dot and on the QPC) therefore increases the phase space for
QPC-assisted cotunneling processes. Hence, the total rate,
$W_{\alpha\alpha'}^1$, increases with the applied bias voltages.


 \begin{figure}
 \centering
 \includegraphics[width=0.9\columnwidth]{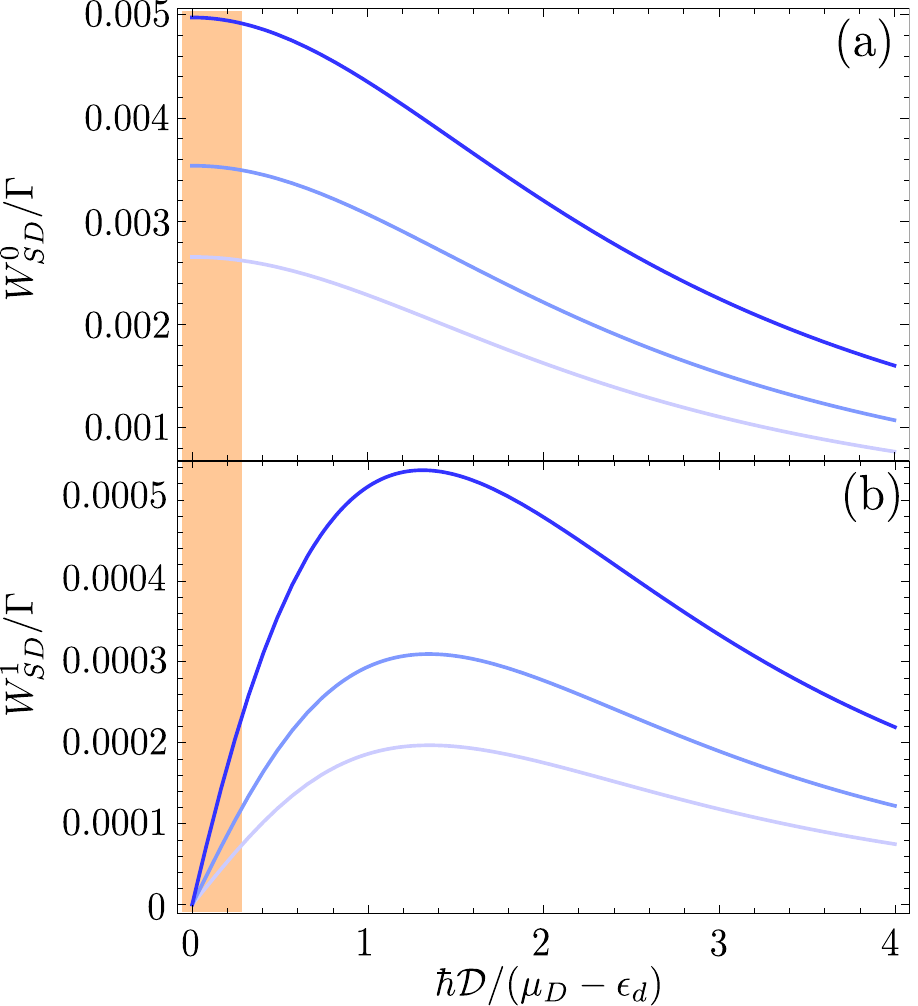}
 \caption[]{\label{Fig:3}
 Plots of (a) $W_{SD}^0$ and (b) $W_{SD}^1$ as a function of $\hbar\mathcal{D}/(\mu_D-\epsilon_d$) with bias voltages $eV_{ SD}=4\hbar\Gamma$, $eV_{ LR}=3.9\hbar\Gamma$, and $\Gamma_S=\Gamma_D$. The different curves are for different dot-gating $\epsilon_d$ such that: $\mu_{ D}-\epsilon_d=4\hbar\Gamma$ (dark blue); $\mu_{ D}-\epsilon_d=5\hbar\Gamma$ (medium-dark blue); $\mu_{ D}-\epsilon_d=6\hbar\Gamma$ (bright blue). We highlight the area in which our toy model is relevant, namely, where $W_{SD}^0$ slightly decreases, $W_{SD}^1$ is far from peaking, and $W_{SD}^n$ for $n>1$ are negligible. Two effects are seen here: (i) The higher the dot-energy is, the higher the cotunneling rates are; (ii) as $\mathcal{D}/(\mu_D-\epsilon_d)$ increases it is more probable for electrons to pass through the QPC during the time-window in which the dot is empty. This is seen in the decrease in $W_{SD}^0$ vs. the increase in $W_{SD}^1$. At the point in which $W_{SD}^1$ peaks, the probability for two electrons to tunnel through the QPC during a cotunneling event becomes relevant.}
 \end{figure}

In Fig.~\ref{Fig:3}, we plot the rates $W_{SD}^0$ and $W_{SD}^1$ as a function of $\hbar\mathcal{D}/(\mu_D-\epsilon_d)$ for different values of $\mu_D-\epsilon_d$ and at fixed $V_{ SD}$, $V_{LR}$. In this configuration, $\mu_D-\epsilon_d$ affects all possible cotunneling rates. Lowering $\epsilon_d$ suppresses the rate of cotunneling processes, leading also to shorter cotunneling times $\tau_{\rm cot}$. This is seen in Fig.~\ref{Fig:3} by the overall decrease in magnitude of both $W_{SD}^0$ and $W_{SD}^1$ as $\epsilon_d$ is lowered. As $\mathcal{D}$ increases the probability for an electron to tunnel through the QPC during the time window $\tau_{\rm cot}$ grows. Hence, $W_{SD}^0$ constantly decreases with $\mathcal{D}$ while $W_{SD}^1$ increases. 

At a certain point, $W_{SD}^1$ peaks and decreases as well. At this point the probability for two electrons to tunnel through the QPC during a cotunneling event becomes relevant. We highlight the regime in which our toy model is valid, i.e.~, the regime in which $W_{SD}^0$ slightly decreases while $W_{SD}^1$ increases but is still far from peaking. Note, however, that even beyond the highlighted regime, in the weak QPC-tunnel-coupling regime the rates $W_{SD}^n$ with $n>1$ appear as contributions of $O(\Omega^{2n})$.

We have assumed throughout this section an additional finite width $\hbar\mathcal{D}/2$ of the dot energy level. We stress that this is a result of the formal derivation of Eqs.~\eqref{eq: Wsd0} and \eqref{integralForm} that appears in Appendix \ref{tunnelRates}.

\subsubsection{Currents and correlations}
Using the rates, Eqs.~\eqref{eq: Wsd0} and \eqref{integralForm}, we are able to express the currents and cross-correlations of our setup~\cite{Korotkov:1994,Koch:2006}. The average currents of the dot and the QPC are given by
\begin{align}\label{dot_current_eq}
&\langle I_{\rm dot}\rangle=e\left(W_{SD}^0+W_{SD}^1-W_{DS}^0-W_{DS}^1\right)~,\\
&\langle I_{\rm QPC}\rangle=e\left(W_{SD}^1+W_{DS}^1+W_{SS}^1+W_{DD}^1\right)~.\label{qpc_current_eq}
\end{align}
The zero-frequency cross-correlation between these currents,
\begin{equation}
\label{correlazioni}
S=2\int_{-\infty}^{\infty}d\tau \left[\langle I_{\rm dot}(t+\tau)I_{\rm QPC}(t)\rangle-\langle I_{\rm dot}\rangle\langle I_{\rm QPC}\rangle\right]~,
\end{equation}
 is given by
\begin{equation}
\label{correlazioni2}
S=2e^2(W_{SD}^1-W_{DS}^1)~.
\end{equation}

\section{Results}
\label{result}
In this section, we analyze the resulting currents through the dot and
the QPC, and their cross-correlations. 
Also, we relate these quantities to a conditional partial measurement of the occupation of the dot, akin to null weak values~\cite{Zilberberg2013}. This quantity enables the determination of $\tau_{\rm cot}$.

\subsection{Current through the dot; measurement back-action}
\begin{figure}
 \centering
 \includegraphics[width=\columnwidth]{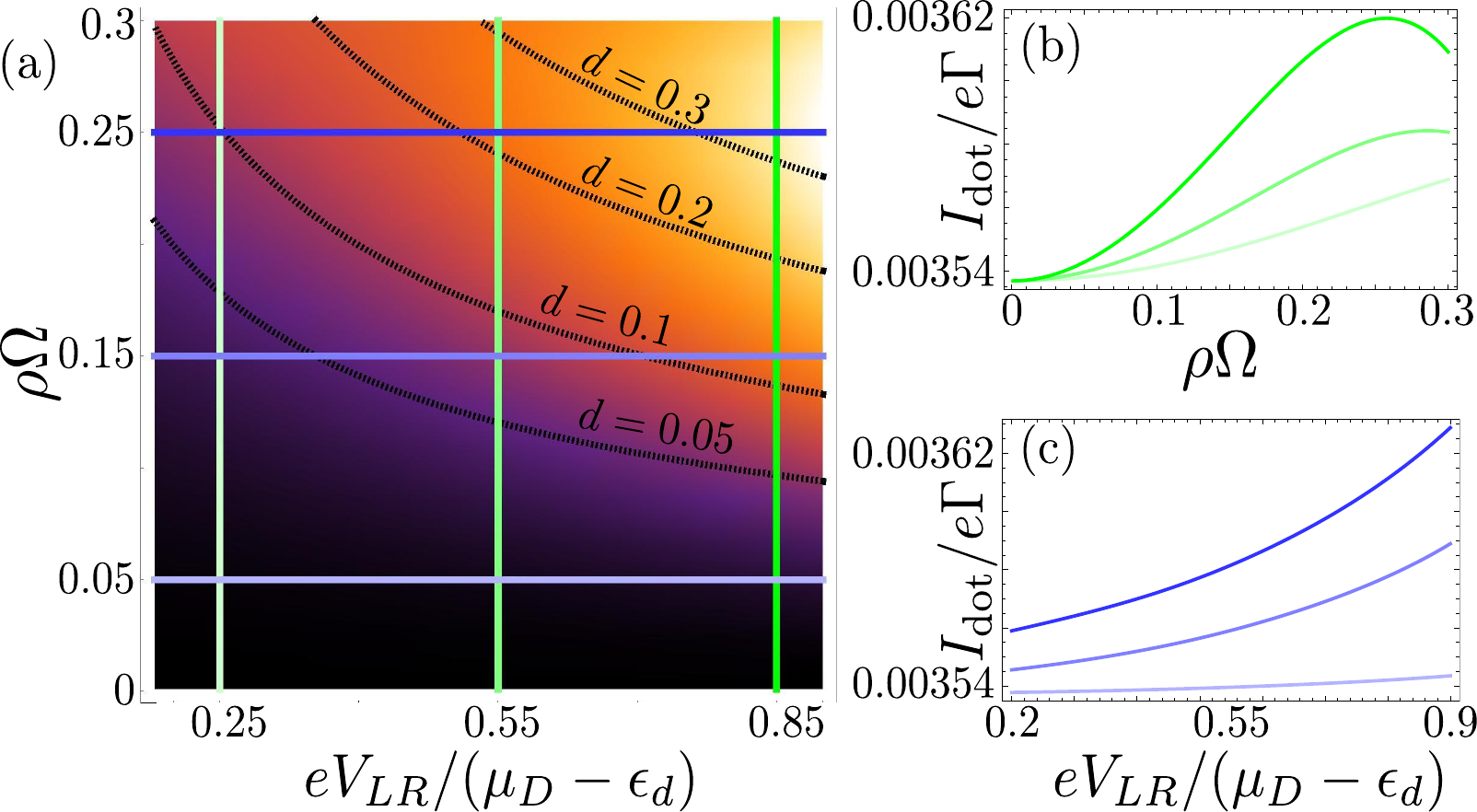}
 \caption[]{\label{Fig:4}
 Plots of $I_{\rm dot}$ [cf.~Eq.~\eqref{dot_current_eq}] as a function of $eV_{LR}/(\mu_D-\epsilon_d$) and $\rho\Omega$ with $eV_{ SD}=4\hbar\Gamma$, $\mu_{ D}-\epsilon_d=5\hbar\Gamma$, and $\Gamma_S=\Gamma_D$. (a) A density plot of $I_{\rm dot}$. The $I_{\rm dot}$ along the vertical (green) and horizontal (blue) mesh lines is plotted in (b) and (c), respectively. The dashed lines represent equal-$d\equiv \hbar\mathcal{D}/(\mu_D - \epsilon_d)$ lines. We can see that the current increases with the coupling $\mathcal{D}$ of the QPC. However, the effect does not depend exclusively on $\mathcal{D}$, but it also directly depends on the QPC parameters: the tunnel-coupling $\Omega$ and the voltage-bias $V_{LR}$.}
 \end{figure}

The current through the dot is carried by cotunneling processes, in
which the occupation of the dot is virtually changed for a short
time. 
The intermediate evolution state, where the occupation of the dot is
changed, does not preserve energy. 
Thus, ideally a strong measurement of the charge on the dot would
destroy cotunneling processes.
Given that the current through the QPC is affected by the charge on
the dot, it can be used to measure this charge. 
Naively, one would expect that turning on a weak coupling between the
QPC and the dot, will slightly reduce the current through the dot. 
It turns out that the opposite is true. 
As seen in Fig.~\ref{Fig:4}, by increasing the coupling between the dot and the QPC (increasing $\Omega$ in our toy model), the current through the dot is even slightly \emph{enhanced}.

Figure~\ref{Fig:4} depicts the current through the dot as a function
of the coupling to the QPC, $\Omega$, and as a function of the QPC
voltage-bias, $V_{LR}$. 
The current is enhanced both by increasing $\Omega$ and by increasing
$V_{LR}$, or alternatively, by increasing $\mathcal{D}$, and
$V_{LR}$. 
Note that the regime in which our model is valid is $d\equiv
\hbar\mathcal{D}/(\mu_D - \epsilon_d)\lesssim 0.3$, where up to one electron is transfered in the QPC during the virtual cotunneling time of the dot.
Therefore, the area in which $I_{\rm dot}$ peaks and decreases is outside the scope of our model's validity.

The current through the dot has two parts, a ``coherent'' part and
``incoherent'' part. 
The coherent part is carried by coherent cutunneling processes, where
electrons tunnel through the dot without changing the state of the
QPC, the total rate of these processes is $W_{SD}^0$. 
The incoherent part is carried by cotunneling processes through the
dot that are accompanied by changes in the state of the QPC, namely,
tunneling through the QPC. 
The rates for these processes are $W_{SD}^1$ and $W_{DS}^1$. 
As discussed in Sec.~\ref{derivationOfRates}, increasing $\Omega$ or
$V_{LR}$, and correspondingly $\mathcal{D}$, increases the probability
for tunneling through the QPC, and hence decreases $W_{SD}^0$ and
increases $W_{SD}^1$. 
Additionally, $V_{LR}$ increases also the phase-space for inelastic
QPC-assisted cotunneling processes through the dot. Hence $V_{LR}$
increases $W_{SD}^1$ (and $W_{DS}^1$) even for a fixed $\mathcal{D}$. 
Hence, the growth of $I_{\rm dot}$ in Fig.~\ref{Fig:4} is a result of the
availability of phase-space for inelastic QPC-assisted cotunneling
processes in $W_{SD}^1$ being higher than the decrease in $W_{SD}^0$
as a function of these parameters.


The strong measurement nature of the incoherent channel is highlighted
by the fact that, in our simplified toy model, having current through
the QPC destroys the coherence of a cotunneling process. 
If the dot would have been embedded into the arm of a Mach-Zehnder
interferometer a current in the QPC could serve as a \emph{strong}
which-path measurement and would reduce the interference signal.  
In this case, the interference signal would be proportional to the
decreasing $W_{SD}^0$. 

\subsection{Current through the QPC; measurement signal}
The current through the QPC is sensitive to the virtual changes in the
charge on the dot during cotunneling processes. 
In the discussed toy model, QPC-current pulses (electron tunneling)
occur only alongside cotunneling processes in the dot. 
The cotunneling processes can be either between two leads (from the
source to the drain and vice versa), or back and forth between the dot
and one of the leads (source-to-source or drain-to-drain). We find, here, a non-monotonous signal that is dependent on the directionality of the dot-transport.

\begin{figure}
 \centering
 \includegraphics[width=\columnwidth]{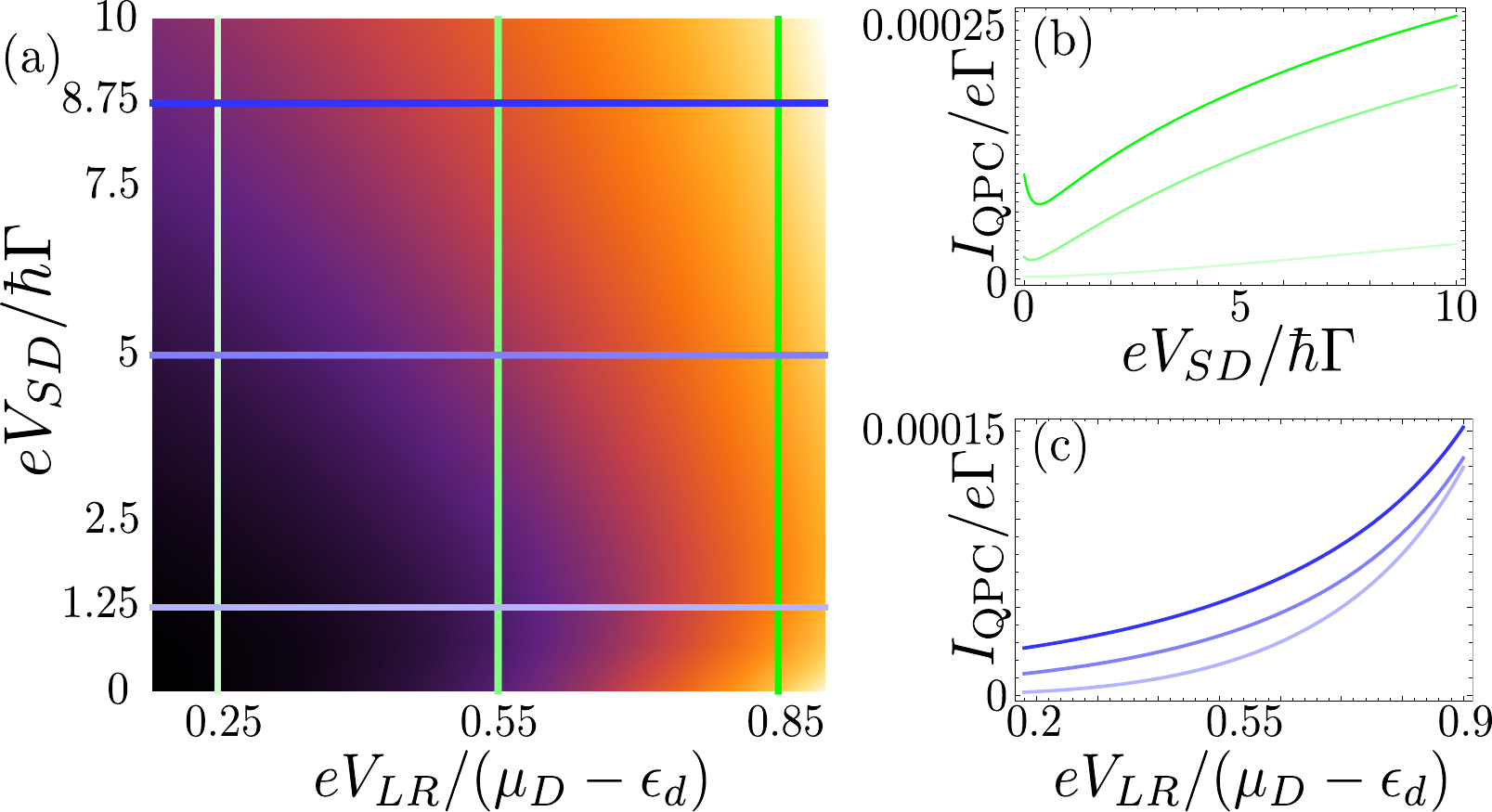}
 \caption[]{\label{Fig:5}
 Plots of $I_{\rm QPC}$ [cf.~Eq.~\eqref{qpc_current_eq}] as a function of $eV_{LR}/(\mu_D-\epsilon_d$) and $eV_{SD}/(\hbar\Gamma)$ with $\rho\Omega=0.15$, $\mu_{ D}-\epsilon_d=5\hbar\Gamma$, and $\Gamma_S=\Gamma_D$. (a) A density plot of $I_{\rm QPC}$. The $I_{\rm QPC}$ along the vertical (green) and horizontal (blue) mesh lines is plotted in (b) and (c), respectively. Naturally, the current increases with the the QPC voltage-bias $V_{LR}$. As the voltage on the dot $V_{SD}$ is increased the following occurs: (i) The cotunneling to the drain and back $W_{DD}^1$ remains constant; (ii) The cotunneling to the source and back $W_{SS}^1$ becomes less probable; (iii) The cotunneling from drain to source $W_{DS}^1$ decreases, and disappears once $V_{SD}>V_{LR}$; (iv) Additional cotunneling processes from source to drain in $W_{SD}^1$ are allowed. The latter features dominate the behavior of $I_{\rm QPC}$. We see initially a slight decrease as $W_{DS}^1$ vanishes with an overall increase due to $W_{SD}^1$. }
 \end{figure}

In Fig.~\ref{Fig:5}, we plot $I_{\rm QPC}$ as a function of the two
bias voltages: $V_{SD}$ on the dot, and $V_{LR}$ on the QPC. 
Increasing $V_{LR}$ has two effects: First, it increases the phase
space for elastic QPC-tunneling processes, where
$\epsilon_l=\epsilon_r$ (see Fig.~\ref{Fig:2} for clarification of the
notations); Second, it increases the phase space for inelastic processes with
$\epsilon_l>\epsilon_r$. 
While the first effect is relevant only for source-to-drain tunneling
through the dot, the second effect is relevant for all four
possibilities of cotunneling through the dot (from any lead $\alpha$
to any lead $\alpha'$). 
The amount of energy that can  be lost in the QPC as a compensation
for inelastic cotunneling through the dot is bound from above by
$eV_{LR}$. Hence, in cotunneling processes that involve a single lead,
raising $V_{LR}$ allows for more energetic particles-hole excitations
to appear in the lead [see Figs.~\ref{Fig:2}(c) and \ref{Fig:2}(d)]. 
Similarly, drain-to-source cotunneling consumes energy and therefore
increasing $V_{LR}$ enlarges the availability of drain-to-source
cotunneling [see Fig.~\ref{Fig:2}(b)]. 
To conclude, the rates $W_{\alpha\alpha'}^1$ (all the four
combinations) increase with $V_{LR}$ and hence $I_{\rm QPC}$ increases
with $V_{LR}$. 

Raising the bias voltage on the dot, $V_{SD}$, increases the phase
space for both elastic and inelastic source-to-drain tunneling, and
hence, it increases the rate $W_{SD}^1$. 
The rate $W_{DS}^1$ however, is suppressed by a raised $V_{SD}$ as the
drain-to-source cotunneling processes require higher amount of energy
for larger $V_{SD}$. 
For a fixed $V_{LR}$, the available energies for inelastic for
drain-to-source cotunneling through the dot are bounded from
above. Hence, increasing $V_{SD}$ reduces the phase space for such
processes, where for $V_{SD}>V_{LR}$ such cotunneling becomes impossible. 

The increase of $V_{SD}$ has an additional effect; for a fixed $\mu_D$
it reduces the rate $W_{SS}^1$. 
The probability for a cotunneling process to occur, depends on the
energy difference between the dot and the available energies in the
lead. Increasing $\mu_S-\epsilon_d$, increases the difference between
the energy of the electron in the dot and the available energies in
the source [see Fig.~\ref{Fig:2}(c)], making the source-to-source
cotunneling processes less probable. 
Since $\mu_D$ is fixed, the drain-to-drain tunneling processes are not
affected by $V_{SD}$ and hence, for a fixed $\mu_D$, $W_{DD}^1$ is
independent of $V_{SD}$.

To conclude, $W_{SD}^1$ grows with $V_{SD}$, while $W_{DS}^1$ and $W_{SS}^1$
are reduced by its increase, and $W_{DD}^1$ is not affected by it. 
As a result, $I_{\rm QPC}$ generally grows with $V_{SD}$ since the
dominant processes are QPC-tunneling processes accompanied by
source-to-drain cotunneling through the dot. 
Yet, for a small bias voltage on the dot, $V_{SD}\ll
\mu_D-\epsilon_d$, the four rates $W_{SD}^1$, $W_{DS}^1$, $W_{SS}^1$,
and $W_{DD}^1$ are roughly of the same magnitude and are determined
mostly by $V_{LR}$. 
Raising $V_{SD}$ in this configuration causes an increase in
$W_{SD}^1$ alongside a decrease in $W_{DS}^1$ and $W_{SS}^1$, causing the
current through the QPC to slightly decrease. 
$I_{\rm QPC}$ starts to increase when $V_{SD}$ is further
increased. This can be shown in Fig.~\ref{Fig:5}(b), where for a
relatively large $V_{LR}$, a small decrease in $I_{\rm QPC}$ appears
as $V_{SD}$ grows, before the overall current starts to increase.

\subsection{Cross-current correlation; focusing on source-drain processes}

We finally arrive to describe the sensing of virtual change in the charge on the dot
during cotunneling processes from source to drain. 
Here, we are interested in the cross-current correlation between the
dot and the QPC. This correlation eliminates processes that contribute to the QPC-current from cotunneling
processes to a specific lead and back, as they do not generate a current through the dot.

\begin{figure}
 \centering
 \includegraphics[width=\columnwidth]{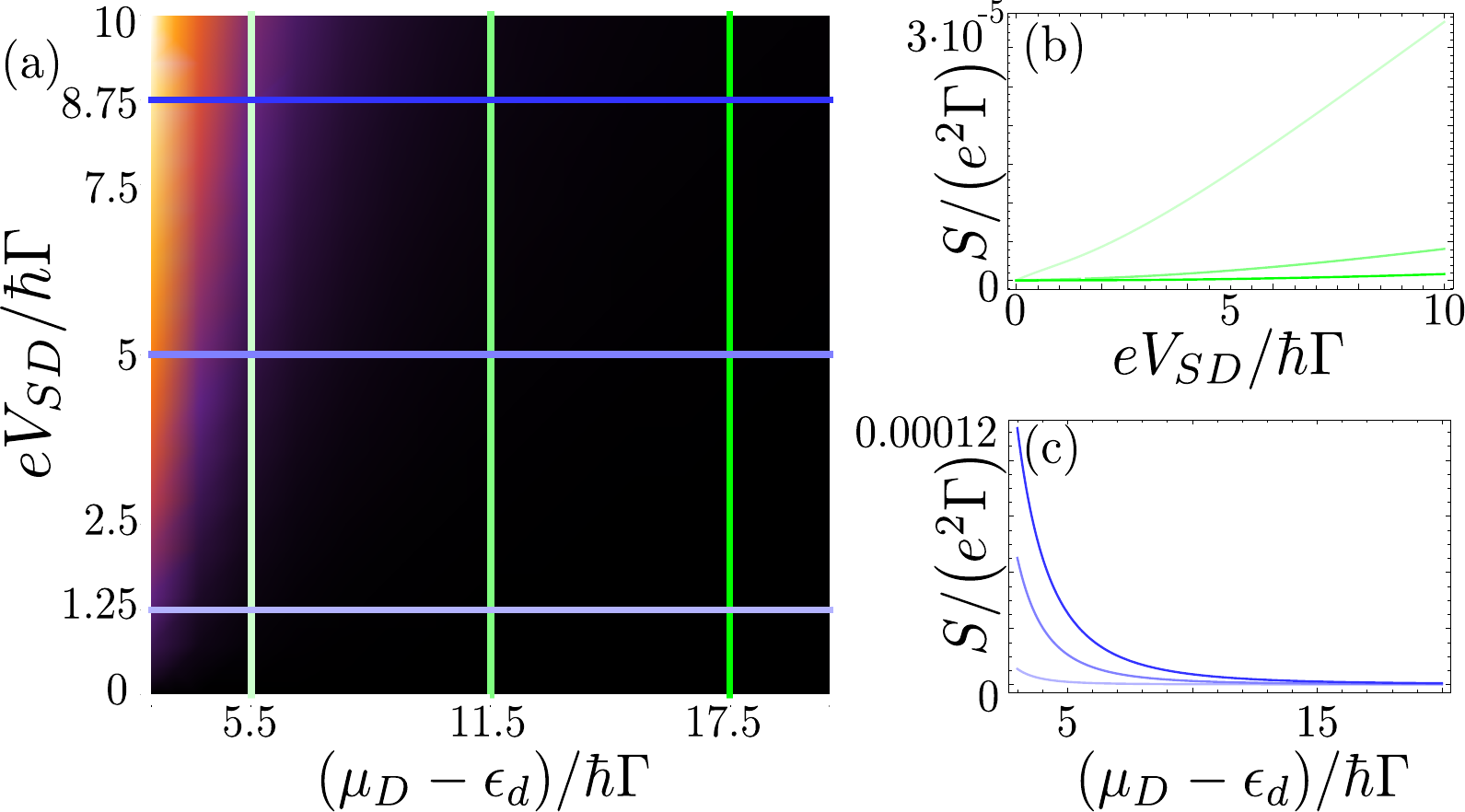}
 \caption[]{\label{Fig:6}
 Plots of $S$ [cf.~Eqs.~\eqref{correlazioni} and \eqref{correlazioni2}] as a function of $eV_{SD}/(\hbar\Gamma)$ and $(\mu_D-\epsilon_d)/(\hbar\Gamma)$ with $\rho\Omega=0.15$, $eV_{LR}=1.5\hbar\Gamma$, and $\Gamma_S=\Gamma_D$. (a) A density plot of $S$. The $S$ along the vertical (green) and horizontal (blue) mesh lines is plotted in (b) and (c), respectively. As the voltage on the dot $V_{SD}$ is increased the following occurs: (i) The cotunneling from drain to source $W_{DS}^1$ decreases, and disappears once $V_{SD}>V_{LR}$; and (ii) additional cotunneling processes from source to drain in $W_{SD}^1$ are allowed. Hence, $S$, initially, increases with $V_{SD}$ slowly, and only afterward increases as $W_{SD}^1$. The deeper the dot level is, the shorter the virtual time in which the dot is empty. As a result, (i) cotunneling processes through the dot become less probable, and (ii) it becomes less probable for a QPC electron to manage to tunnel during this time window, leading to the descent of $S$. }
\end{figure}

The correlation function, $S$, in Eq.~(\ref{correlazioni}) is plotted in Fig.~\ref{Fig:6}, as a function of the system's parameters, $V_{SD}$ and $\mu_D-\epsilon_d$. 
Its behavior stems from the same effects considered above for the currents, $I_\textrm{dot}$ and $I_{QPC}$. 
The correlation is suppressed at $V_{SD}\ll V_{LR}$ because of counter-propagating QPC-assisted processes. Once the drain-to-source processes ($W_{DS}^1$) are suppressed for $V_{SD}\geq V_{LR}$, the correlation increases monotonically with $V_{SD}$ because of the increased phase space for cotunneling events. Also, quite intuitively, the deeper the dot's energy level is, the smaller is the probability for a cotunneling event to occur, and thus the smaller $S$ becomes. As to the effects of the detector on the correlation, that we do not plot, a larger coupling to the QPC or a larger $V_{LR} $ leads to a stronger current across the detector, which directly reflects in a larger signal in the correlation.

\subsection{Weak values; cotunneling time}
\begin{figure}
 \centering
 \includegraphics[width=\columnwidth]{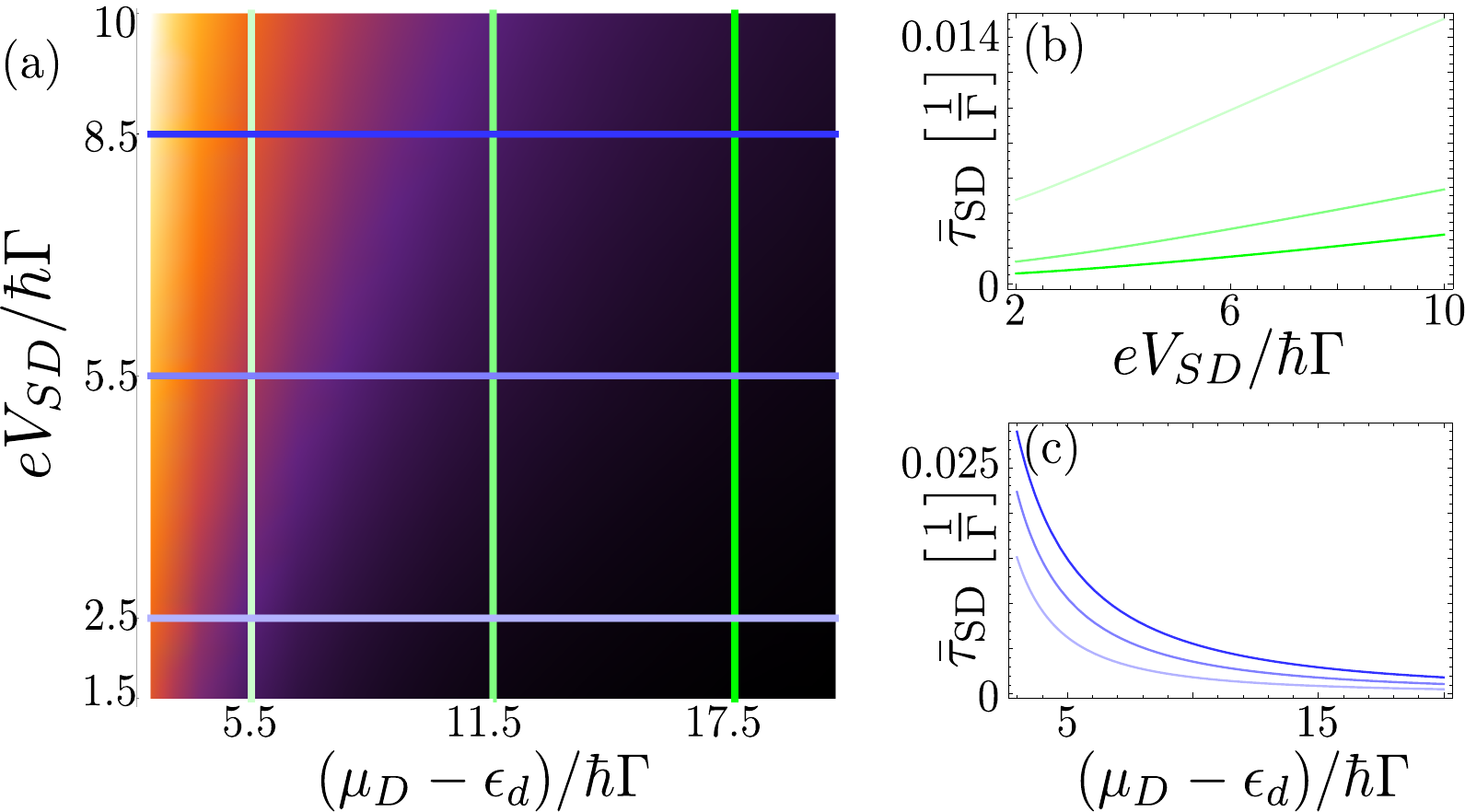}
 \caption[]{\label{Fig:7}
 Plots of $\bar{\tau}_{\textrm{SD}}$ [cf.~Eq.~\eqref{tau}] as a function of $eV_{SD}/(\hbar\Gamma)$ and $(\mu_D-\epsilon_d)/(\hbar\Gamma)$ with $\rho\Omega=0.15$, $eV_{LR}=1.5\hbar\Gamma$, and $\Gamma_S=\Gamma_D$. (a) A density plot of $\bar{\tau}_{\textrm{SD}}$. The $\bar{\tau}_{\textrm{SD}}$ along the vertical (green) and horizontal (blue) mesh lines is plotted in (b) and (c), respectively. Interestingly, as $V_{SD}$ increases, $\bar{\tau}_{\textrm{SD}}$ increases as well, i.e. it appears that the increased phase-space adds processes with a slower time into the average. As expected, the deeper the dot level is, the shorter $\bar{\tau}_{\textrm{SD}}$ becomes. }
\end{figure}

The current through the QPC is generated while the dot is virtually empty, i.e.~$I_{\textrm{QPC}}=\sum_i I_{\textrm{QPC}}^{\{i\}}$ is generated by processes $i$ for which the dot is virtually empty for time $\tau_{\rm cot}^{\{i\}}$ with $I_{\textrm{QPC}}^{\{i\}}\sim e\mathcal{D}\tau_{\rm cot}^{\{i\}}$. 
Since the correlation function, $S$, isolates the contributions to the QPC current arising from cotunneling between the two distinct leads, it encodes information on the physical properties of these cotunneling processes, i.e.~one can extract the time the dot is virtually empty, when restricting to current-generating cotunneling events. 
This can be defined via a weak value procedure~\cite{Aharonov:1988a}, where the detector's signal is postselected by retaining only processes of successful cotunneling current from source to drain. 
In this framework the cotunneling current through the dot plays the role of a postselection operator~\cite{Romito2014}. 
This leads to the average over cotunneling times of current-generating events~\cite{Romito2014} [cf.~Eq.~(\ref{correlazioni})]
 \begin{align}
 \bar{\tau}_{\textrm{SD}} &=  
\frac{ \int_{- \infty} ^{\infty}ds  [ \langle  I_{\textrm{QD}}
  (t)  I_{\textrm{QPC}} (t-s)   \rangle - \langle I_{\textrm{QD}}
  \rangle \langle I_{\textrm{QPC}} \rangle]}{e\mathcal{D} \langle I_{\textrm{QD}}\rangle}\nonumber\\
	&=\frac{S}{2e\mathcal{D} \langle I_{\textrm{QD}}\rangle}\,.
	\label{tau}
 \end{align}
The obtained $\tau_\textrm{SD}$ is plotted in Fig.~\ref{Fig:7} as a function of $eV_{SD}$ and $\mu - \epsilon_d/(\hbar\Gamma)$. 

We note that such a quantity generally depends on the detector's parameters as well. 
However, the effects of the detector are minimal in the weak measurement regime, which, as discussed in Sec.~\ref{method}, corresponds to $\hbar \mathcal{D}\ll eV_{LR} \ll \mu_D -\epsilon_d$. This allows us to define an intrinsic (dot-dependent only) cotunneling time when $V_{LR} \to 0$. The evaluation of $\bar{\tau}_\textrm{SD}$ that is reported in Fig.~\ref{Fig:7} is obtained  within such a regime, and its values do not depend on the detector parameters much. 

In Fig.~\ref{Fig:7}, we see that $\bar{\tau}_{\textrm{SD}}$ increases with $V_{SD}$, and becomes shorter the lower the dot-level is. The decay of the cotunneling time as a function of $\mu_D-\epsilon_d$ can be intuitively understood in terms of the energy-time uncertainty principle: at equilibrium, the virtual hole that is excited in the dot sets $\hbar/ (\mu_D-\epsilon_d)$ as the typical time scale for the occupation of the virtual state.

\subsubsection{Discussion}
\label{discussion}
We wish to make a more quantitative evaluation of our obtained cotunneling time [cf.~$\bar{\tau}_{SD}$ in Eq.~\eqref{tau} and Fig.~\ref{Fig:7}]. Due to the finite bias voltage on the dot, in Eq.~\eqref{tau} we have a weighted average over the cotunneling times of different cotunneling processes. Within a simple comparative model, this can be taken into account by averaging the cotunneling times predicted by Heisenberg's uncertainty principle, $\tau_h=\hbar/ (\epsilon_h-\epsilon_d)$, over all cotunneling processes, each characterized by a different virtual energy, $\epsilon_h-\epsilon_d$, where $\mu_D< \epsilon_h < \mu_S$. Each cotunneling process occurs with probability $P(\epsilon')=2\pi\rho_S \rho_D \left|t_St_D/(\epsilon'-\epsilon_d-i\hbar\mathcal{D}/2)\right|^2$ [cf.~Eq.~\eqref{eq: Wsd0}]. The averaged cotunneling time is then
\begin{widetext}\begin{align}
\bar{\tau}_h=\frac{\int_{\mu_D}^{\mu_S} P(\epsilon') \tau_h(\epsilon')}{\int_{\mu_D}^{\mu_S} P(\epsilon')}=\frac{1}{\hbar \mathcal{D}}\frac{\ln\left[(\hbar \mathcal{D})^2 + 4 (\mu_S-\epsilon_d)^2\right]-\ln\left[(\hbar \mathcal{D})^2 + 4 (\mu_D-\epsilon_d)^2\right] + 2 \ln[\mu_D-\epsilon_d]- 2 \ln[\mu_S-\epsilon_d]}{ \arctan\left[2 (\mu_S-\epsilon_d)/(\hbar \mathcal{D})\right]+\arctan\left[2 (\mu_D-\epsilon_d)/(\hbar \mathcal{D})\right]}\,.
\label{tauEff}
\end{align}\end{widetext}

\begin{figure}
 \centering
 \includegraphics[width=\columnwidth]{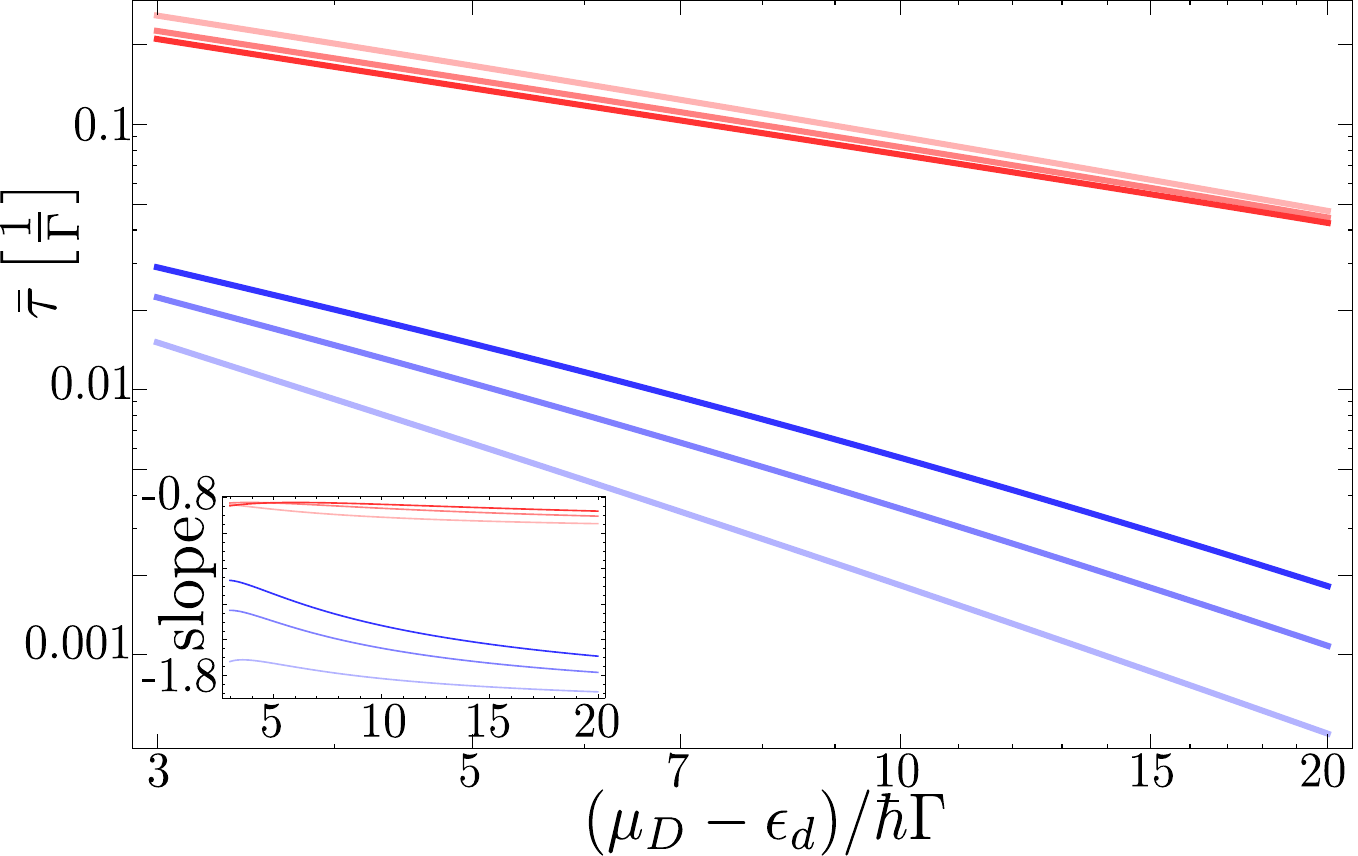}
 \caption[]{\label{Fig:8}
 LogLog plots of $\bar{\tau}_{\textrm{SD}}$ and $\bar{\tau}_{h}$ [cf.~Eqs.~\eqref{tau} and \eqref{tauEff}] as a function of $(\mu_D-\epsilon_d)/(\hbar\Gamma)$ with $\rho\Omega=0.15$, $eV_{LR}=1.5\hbar\Gamma$, and $\Gamma_S=\Gamma_D$. The plots of $\bar{\tau}_{\textrm{SD}}$ have an overall shorter time (blue) than those of $\bar{\tau}_{h}$ (red). For both plotted times the curves are for $V_{SD}=2.5 \hbar \Gamma$ (light), $V_{SD}=5.5 \hbar \Gamma$ (medium-light), and $V_{SD}=8.5 \hbar \Gamma$ (dark). In the inset, we plot the approximate slopes of the LogLog curves, defined as $(\partial \ln\bar{\tau})/[\partial \ln(\mu_D-\epsilon_d)]$. The overall magnitude difference between the two time models can be attributed to an unknown prefactor in the time taken from the energy-time uncertainty principle. The other differences can be attributed to an inherent discrepancy between the two  models, as the former incorporates phase-space contributions of slower QPC-assisted processes.}
\end{figure}

The resulting $\bar{\tau}_h$ is plotted in Fig.~\ref{Fig:8}, where it is compared to $\bar{\tau}_\textrm{SD}$. 
We see the following main differences: (i) there is, approximately, an overall order of magnitude between the two times, which can be attributed to an unknown prefactor in the used $\tau_h (\epsilon')$ from the energy-time uncertainty principle, (ii) whereas $\bar{\tau}_h$ decreases almost as $1/(\mu_D-\epsilon_d)$ (it exhibits almost a linear decrease in a LogLog plot) with a slight dependence on $V_{SD}$, the $\bar{\tau}_\textrm{SD}$ deviates from such a $1/(\mu_D-\epsilon_d)$ descent and shows a stronger dependence on $V_{SD}$, (iii) the dependence on $V_{SD}$ is opposite in the two cases. With increasing $V_{SD}$, in $\bar{\tau}_h$ additional faster cotunneling processes are added into the average and it becomes shorter, whereas in $\bar{\tau}_\textrm{SD}$ the average time becomes longer. The latter two differences demonstrate the inherent discrepancy between our simplistic model for  $\bar{\tau}_h$, that takes into account only the finite width of the dot level, versus the physical model of $\bar{\tau}_\textrm{SD}$ that incorporates also an increased back-action from the detector by adding phase-space for slower QPC-assisted processes.


\section{Conclusions and Outlook}
\label{conclude}
We have presented a model for a direct detection of
electron cotunneling through a single-level quantum dot. 
The detector has been modeled by 
a charge sensing QPC that is capacitively coupled to the dot.
In the regime where the transport through the QPC
is fully blocked when the dot is occupied, we obtain a simplified model 
 that allows us to incorporate a microscopic description of the QPC into the rate equation formalism of
cotunneling through the quantum dot.
We have, thus, determined the current through the QPC (detector's signal), the cotunneling current in the dot (including the detector's back-action) as well as their correlations (related to weak values and the cotunneling time).

We found that the detector's back-action consists of three different mechanisms: (i) a broadening of the dot energy levels, (ii) a suppression of elastic (coherent) charge transfer processes, and (iii) an increase of phase-space due to QPC-assisted transport. In particular, the latter mechanism is responsible for an increase of the cotunneling current upon increasing the QPC voltage bias. The QPC current generally increases also as a function of the dot voltage bias corresponding to the increased probability of cotunneling processes. However, a counter-intuitive \emph{decrease} of $I_\textrm{QPC}$ with the voltage bias across the dot is predicted at small voltage biases. In this regime, the dominant effect of the increase of voltage bias is a suppression of the source-to-source, drain-to-drain, and drain-to-source cotunneling processes, which in turn result in a suppressed current through the QPC. Importantly, the current-current correlations allow us, via a weak value based approach, to access the cotunneling time. We find that the cotunneling time obtained from such a direct measurement can be compared to the time estimated from the energy-time uncertainty principle, after taking into account proper averaging of all possible cotunneling events. 

Let us, finally, comment on the validity and applicability of our simplified model. 
Our model essentially relies on the assumptions that (i)  transport through the QPC is fully blocked for an occupied dot, and that (ii) only processes of single electron transfer through the QPC  during a cotunneling event are considered. The former assumption is used in order to obtain time-independent cotunneling rates in the presence of the QPC. Introducing a Markovian thermalization time in the QPC leads or cutting the infinite time integration in the cotunneling rates by sequential tunneling rates should cure this constraint. 
The latter condition appears to be quite strong: in order to detect a cotunneling event of duration $\tau_\textrm{cot}\sim \hbar/(\mu_D-\epsilon_d)$, the detector is expected to  have a large bandwidth $\gtrsim \hbar/\tau_{\rm cot}$, corresponding to a large number of charge-transfer events in the QPC per single cotunneling event~\cite{Romito2014}. 
In our model we consider the opposite regime. Nonetheless, the detector is equally sensitive to the short-lived virtual cotunneling states, at the working point defined by (i), since cotunneling processes are the only possible mechanism of activating a current signal in the QPC. This implies that any weak signal in the detector corresponds to obtaining full information on the happening of cotunneling, which makes our measurement of a partial-collapse type (i.e., a strong measurement that happens with a small probability). Hence, unlike in a weak measurement case,  the measured cotunneling time is affected by back-action.

Even in its simplified form, it is important to stress that our simplified theoretical model provides a valid description for experiments at the pinched-off working point of the QPC. 
In this working point, the only approximation is the assumption that the relevant rates are those corresponding to single electron transfer across the QPC. 
Though this is the key simplification, it is also a realistic physical approximation. 
What our approximation neglects are the coherences between subsequent electrons tunneling across the QPC, which decay very fast due to the relaxation processes of electrons in the bulk leads. Hence, contributions from a higher number of transported QPC-electrons per cotunneling process are small (cf.~Fig.~\ref{Fig:3}). In other words, the virtual cotunneling time is so short that the assumption $\mathcal{D}\tau_{\rm cot}\ll 1$ is physically sensible.

\textit{Spinful electrons.} In order to highlight the interplay between measurement and coherent transport, we have assumed that the electrons are spin
less. Whereas spin physics do not play an important role in QPC
transport, in the dot system the Kondo effect~\cite{hewson1997kondo} could
qualitatively change our predictions. 
Our treatment is immediately applicable for polarized electrons, e.g. in
the presence of a strong magnetic field, which could be realized in
experiments. Moreover, also in the case of complete spin degeneracy,
where Kondo physics takes place, our approach remains valid as long as
the Kondo temperature $T_k \sim \sqrt{\rho_\alpha |t_\alpha|^2 U}
e^{-U/(2\rho_\alpha |t_\alpha|^2)}$, with $U$ the charging energy on
the dot, is smaller than the different energy scales of the system,
e.g.~, $T_K \ll T\ll eV_{SD}$. In this regime, the system is not
sensitive to the Kondo physics and our treatment is essentially
correct. Note that this specific limit can be achieved by making the
tunneling between leads and dot arbitrarily weak. Importantly,
extending our analysis to the Kondo regime is extremely interesting,
as it features an interplay between transport through a truly
many-body correlated virtual state and weak measurement.

Since our model can describe realistic configurations~\cite{Field1993,Elzerman2003,DiCarlo2004,Harbusch2010,Gasparinetti2012,Granger2012,clemens2012,clemens2013,zumbuhl2014}, it becomes interesting to generalize it to include experimentally relevant effects, e.g. finite temperature, many-level dots, spin physics, and to extend the present results to the more general regime of weak measurement.

\acknowledgments 
 We would like to thank S. A. Gurvitz, Y. Oreg, T. Ihn, C. R\"{o}ssler, K. Ensslin, and D. Z\"{u}mbuhl for useful discussions. We acknowledge the support from DFG, 4710/1-1, and Swiss
National Science Foundation (SNSF).

\appendix
\section{Derivation of the cotunneling rates in the presence of coupled QPC}
\label{rigorous}
We present here a rigorous model for the calculation of the cotunneling rates appearing in Eqs.~\eqref{eq: Wsd0} and \eqref{integralForm}. We start by presenting a general model for the analysis of a QPC current in Appendix \ref{qpcAmp}. Using this formalism we derive the rates in Appendix \ref{tunnelRates}.

\subsection{QPC analysis}
\label{qpcAmp}
We present here the microscopic derivation of the tunneling probability amplitudes describing
electron transport in the QPC. This derivation is closely related to the one presented in Ref.~\onlinecite{gurvitz}.

The QPC is out-of-equilibrium, namely, the chemical potentials of the left and right leads are different, $\mu_L=eV_{LR}/2$, $\mu_R=-eV_{LR}/2$. Hence, the electron creation and annihilation operators are defined with respect to a vacuum state $\ket{0\; ; \; e}$ in which all the levels in the left (right) lead are initially filled up to
the Fermi energy $\mu_L$ ($\mu_R$). We assumed here that the dot is initially empty, denoted by $e$. Therefore, the vacuum state $\ket{0\; ; \; e}$ decays exponentially to states
with transmitted electrons from the left to the right leads, (e.g.~, $a_r^\dagger a_l\ket{0 \; ; \; e}$, with an electron in the
right lead and a hole in the left lead,
$a_r^\dagger a_{r^\prime}^\dagger a_l a_{l^\prime}\ket{0\; ; \; e}$ with two electrons in the right lead
and two holes in the left lead, etc.).

In the absence of interaction with the dot, the many-body wave function that describes the electron dynamics in the QPC can be
written in the following form
 \begin{align}
 \label{emptyWF}
	\ket{\Psi(t);e}&=\Big[ b_0(t)+\sum_{l_1,r_1}b_{l_1r_1}(t) a_{r_1}^\dagger a_{l_1}\\
&+\sum_{l_1<l_2,r_1<r_2}b_{l_1l_2 r_1r_2}(t) a_{r_1}^\dagger a_{r_2}^\dagger a_{l_1} a_{l_2}+...\Big]\ket{0\, ; e}\, ,\nonumber
\end{align}
where $b_{\mathbf{l}_i,\mathbf{r}_i}(t)$ with $\mathbf{l}_i,\mathbf{r}_i=l_1,...,l_i,r_1,...,r_i$, are the time-dependent probability amplitudes to
find the system in a state with $i$ electrons transmitted from left to right. The indices $l_i,r_i$ label the momenta of the transmitted electrons. The initial
condition is usually chosen to be $b_{\mathbf{l}_i,\mathbf{r}_i}(0)=\delta_{i0}$, and the probabilities are summed to one,
 \begin{align}
	\sum_i\sum_{\mathbf{l}_i,\mathbf{r}_i} \left|b_{\mathbf{l}_i,\mathbf{r}_i}(t)\right|^2=1\, .
\end{align}

The QPC is capacitively coupled to the quantum dot, i.e.~the tunneling amplitude, $\Omega$, is modulated by the charge on the dot. Hence, the wave function of the electrons in the QPC is also modulated by the charge on the dot.

We denote the vacuum state (of the QPC) in the presence of an occupied dot with $\ket{0\; ; \; f}$. Hence, the many-body wave function of Eq.~\eqref{emptyWF} becomes
\begin{align}
\label{fullWF}
	\ket{\tilde\Psi(t);f}&=\Big[ \tilde b_0(t)+\sum_{l_1,r_1}\tilde b_{l_1r_1}(t) a_{r_1}^\dagger a_{l_1}\\
&+\sum_{l_1<l_2,r_1<r_2} \tilde b_{l_1l_2 r_1r_2}(t) a_{r_1}^\dagger a_{r_2}^\dagger a_{l_1} a_{l_2}+...\Big]\ket{0\, ; f},\nonumber
\end{align}
where $\tilde b_{\mathbf{l}_i,\mathbf{r}_i}(t)$ are the time-dependent probability amplitudes in the presence of an occupied dot. These amplitudes satisfy the same normalization condition as the amplitudes $b_{\mathbf{l}_i,\mathbf{r}_i}(t)$.

The explicit expressions for $b_{\mathbf{l}_i,\mathbf{r}_i}(t)$ [$\tilde b_{\mathbf{l}_i,\mathbf{r}_i}(t)$] are obtained by substituting Eq.~(\ref{emptyWF}) [Eq.~(\ref{fullWF})] into the Schr\"odinger equation
$i\hbar|\dot\Psi (t)\rangle ={ H}_{\text{QPC}}|\Psi (t)\rangle$ and choosing appropriate boundary conditions. Note that the time evolution of Eq.~\eqref{emptyWF} is performed by ${ H}_{\text{QPC}}=H_{\text{det}}$, and the evolution in time of Eq.~\eqref{fullWF} is performed by ${ H}_{\text{QPC}}=H_{\text{det}}+H_{\text{int}}$. Therefore, the form of the coefficients $\tilde b(t)$ is the same as that of $b(t)$ with $\Omega$ replaced by $\tilde\Omega\equiv\Omega+\delta\Omega$. Thus, the QPC current can be used to measure the charge of the dot. The strength of the measurement is encoded in the tunneling amplitudes, $\Omega$ and $\tilde\Omega$. In the following it suffices to present the formalism for the case of an empty dot.

We substitute Eq.~(\ref{emptyWF}) into the Schr\"odinger equation. Performing the Laplace transform
\begin{equation}
\label{rig1} {B}(\mathcal{E})=\frac{1}{\hbar}\int_0^{\infty}e^{i\frac{\mathcal{E}t}{\hbar}}b(t)dt\, ,
\end{equation}
yields coupled equations for $B(\mathcal{E})$:
\begin{align}
 \label{rig2} &\mathcal{E} {B}_{0}(\mathcal{E}) - \sum_{l_1 r_1} \Omega {B}_{l_1 r_1}(\mathcal{E})=ib_0(0)\, ,\\
\label{rig3} (&\mathcal{E} + \mathcal{E}_{l_1} - \mathcal{E}_{r_1}) {B}_{l_1 r_1}(\mathcal{E}) - \Omega{B}_0(\mathcal{E})\\
&\quad\quad -\sum_{l_2 r_2}\Omega {B}_{l_1 l_2 r_1 r_2}(\mathcal{E})=ib_{l_1 r_1}(0)\, ,\nonumber\\
\label{rig4}(&\mathcal{E} + \mathcal{E}_{l_1}+\mathcal{E}_{l_2} - \mathcal{E}_{r_1}-\mathcal{E}_{r_2}) {B}_{l_1 l_2 r_1 r_2}(\mathcal{E})
- \Omega {B}_{l_1 r_2}(\mathcal{E})\nonumber\\
&+\Omega {B}_{l_2 r_2}(\mathcal{E}) -\sum_{l_3 r_3}\Omega {B}_{l_1 l_2 l_3 r_1 r_2 r_3}(\mathcal{E})=ib_{l_1 l_2 r_1 r_2}(0)\, ,\\
 &\cdots\cdots\cdots \, ,
\nonumber
\end{align}
where $b_{\mathbf{l}_i \mathbf{r}_i}(0)$ are the initial conditions.

Now, a recursive replacement of each of the amplitudes $B$ in the sum, $\sum\Omega B$, 
by its expression from the subsequent equation, e.g., 
plugging ${B}_{l_1 r_1}(\mathcal{E})$ from Eq.~(\ref{rig3}) into Eq.~(\ref{rig2}) yields
\begin{widetext}
\begin{equation}
\label{rig5}\left [ \mathcal{E} - \sum_{l_1 r_1}\frac{\Omega^2}{\mathcal{E} + \mathcal{E}_{l_1} - \mathcal{E}_{r_1}}
    \right ] {B}_{0}(\mathcal{E}) - \sum_{l_1 l_2 r_1 r_2}
    \frac{\Omega^2}{\mathcal{E} + \mathcal{E}_{l_1} - \mathcal{E}_{r_1}}{B}_{l_1 l_2 r_1 r_2}(\mathcal{E})=ib_0(0)+i\Omega\sum_{l_1 r_1}\frac{b_{l_1 r_1 }(0)}{\mathcal{E} + \mathcal{E}_{l_1} - \mathcal{E}_{r_1}} \,
\end{equation}
\end{widetext}
We replace sums by integrals,
$\sum_{l,r}\rightarrow\int \rho_{L}(\mathcal{E}_{l})\rho_{R}(\mathcal{E}_{r})\,d\mathcal{E}_{l}d\mathcal{E}_r$ ,
where $\rho_{L}$ and $\rho_{R}$ are the density of states in the left and right leads, respectively.
Consequently, the first sum in Eq.~(\ref{rig5}) has two contributions: (i) a sum over the singular part that yields $i\pi\Omega^2\rho_L\rho_R V_{LR}$,
and (ii) as sum over the principal value that usually cancels out (unless the pole is close to the Fermi energy which causes a rescaling of the energy levels). Usually, the second sum in Eq.~(\ref{rig5})
can be neglected. Indeed, the probability amplitudes $B(\mathcal{E})$ take the form of Green's functions. Hence, replacing ${B}_{l_1 l_2 r_1 r_2}(\mathcal{E})\equiv {B} (\mathcal{E},\mathcal{E}_{l_1},\mathcal{E}_{l_2},\mathcal{E}_{r_1},\mathcal{E}_{r_2})\sim 1/(\mathcal{E}+\mathcal{E}_{l_1}+\mathcal{E}_{l_2}-\mathcal{E}_{r_1}-\mathcal{E}_{r_2})$,
the corresponding integral vanishes, provided that
$V_{LR}\gg \Omega^2\rho$. This is true up to corrections due to the initial conditions. In fact, ${B} (\mathcal{E},\mathcal{E}_{l_1},\mathcal{E}_{l_2},\mathcal{E}_{r_1},\mathcal{E}_{r_2})\sim 1/(\mathcal{E}+\mathcal{E}_{l_1}+\mathcal{E}_{l_2}-\mathcal{E}_{r_1}-\mathcal{E}_{r_2})+b_{l_1 l_2 r_1 r_2}(0)$. If there is a non-vanishing initial condition on $b_{l_1 l_2 r_1 r_2}(0)\neq 0$, we replace
the amplitude ${B}_{l_1 l_2 r_1 r_2}(\mathcal{E})$ by
its expression obtained from the subsequent equation. The integral remains vanishing but we take into account the possible initial conditions that may appear in the other coupled equations.

Repeating this procedure for the other equations gives
\begin{widetext}
\begin{align}
 \label{rig6}(\mathcal{E} + i\hbar\mathcal{D}/2) {B}_{0}(\mathcal{E})=&i \sum_{j=0}\left[ \Omega^{j}\sum_{\mathbf{l}_j \mathbf{r}_j}  \frac{b_{\mathbf{l}_j \mathbf{r}_j}(0)}{\prod_{k=1}^{j}[ \mathcal{E} +\sum_{m=1}^{k} (\mathcal{E}_{l_m} - \mathcal{E}_{r_m})]}\right]\, ,\\
\label{rig7} (\mathcal{E} + \mathcal{E}_{l_1} - \mathcal{E}_{r_1} + i\hbar\mathcal{D}/2) {B}_{l_1 r_1}(\mathcal{E})
      - \Omega{B}_{0}(\mathcal{E})=&i \sum_{j=1}\left[ \Omega^{j-1}\sum_{\mathbf{l}_j \mathbf{r}_j}  \frac{b_{\mathbf{l}_j \mathbf{r}_j}(0)\delta_{l_1\mathbf{l}_j^1} \delta_{r_1 \mathbf{r}_j^1}}{\prod_{k=2}^{j}[ \mathcal{E} +\sum_{m=1}^{k} (\mathcal{E}_{l_m} - \mathcal{E}_{r_m})]}\right]\, ,\\
 \label{rig8}(\mathcal{E} + \mathcal{E}_{l_1}+ \mathcal{E}_{l_2} - \mathcal{E}_{r_1} - \mathcal{E}_{r_2} + i\hbar\mathcal{D}/2) {B}_{l_1 l_2 r_1 r_2}(\mathcal{E})& -
      \Omega {B}_{l_1 r_1}(\mathcal{E})+\Omega {B}_{l_2 r_2 }(\mathcal{E})=\nonumber\\
    &  i \sum_{j=2}\left[ \Omega^{j-2}\sum_{\mathbf{l}_j \mathbf{r}_j}  \frac{b_{\mathbf{l}_j \mathbf{r}_j}(0)\delta_{l_1\mathbf{l}_j^1} \delta_{r_1 \mathbf{r}_j^1}\delta_{l_2\mathbf{l}_j^2} \delta_{r_2 \mathbf{r}_j^2}}{\prod_{k=3}^{j}[ \mathcal{E} +\sum_{m=1}^{k} (\mathcal{E}_{l_m} - \mathcal{E}_{r_m})]}\right]\, ,\\
\cdots\cdots\cdots & ,
\nonumber
\end{align}
\end{widetext}
where $\mathcal{D}=2\pi\Omega^2\rho_L\rho_R eV_{LR}/\hbar$.

Performing the inverse Laplace transform
\begin{equation}
b(t)=\frac{1}{2\pi}\int_{-\infty}^{\infty}e^{-i\frac{\mathcal{E}t}{\hbar}}B(\mathcal{E})d\mathcal{E}\, ,
\label{rig9}
\end{equation}
we can incrementally obtain the time-dependent probability amplitudes. The solutions depend on the initial conditions. Let us first show the solution taking the standard initial condition $b_{\mathbf{l}_i\mathbf{r}_i}(0)=\delta_{i0}$
\begin{align}
 \label{rig10}{b}_{0}(t)=&e^{-\frac{\mathcal{D}}{2}t} & \, ,\\
 \label{rig11}{b}_{l_1 r_1}(t)=&\frac{\Omega }{\mathcal{E}_{l_1}-\mathcal{E}_{r_1}}e^{-\frac{\mathcal{D}}{2}t}\left[1-e^{i\frac{(\mathcal{E}_{l_1}-\mathcal{E}_{r_1})t}{\hbar}}\right]& \, ,\\
\cdots\cdots\cdots & & \, .
\nonumber
\end{align}

Taking, for example, the initial conditions $b_{\mathbf{l}_i\mathbf{r}_i}(0)=\delta_{i1}\delta_{\mathbf{l}_i \mathbf{l}_i^0}\delta_{\mathbf{r}_i \mathbf{r}_i^0}$, we obtain
\begin{align}
 \label{rig12} {b}_{0|l_1^0 r_1^0}(t)=&\frac{\Omega }{\mathcal{E}_{l_1^0}-\mathcal{E}_{r_1^0}-i\frac{\mathcal{D}}{2}}e^{-\frac{\mathcal{D}}{2}t}\times\nonumber\\
 &\left[1-e^{i\frac{\left(\mathcal{E}_{l_1^0}-\mathcal{E}_{r_1^0}-i\frac{\mathcal{D}}{2}\right)t}{\hbar}}\right]  \, ,\\
\label{rig13} {b}_{l_1 r_1|l_1^0 r_1^0}(t)=&e^{-\frac{\mathcal{D}}{2}t}e^{i\frac{\left(\mathcal{E}_{l_1}-\mathcal{E}_{r_1}\right)t}{\hbar}}\delta_{l_1 l_1^0}\delta_{r_1 r_1^0}+ O(\Omega^2) \, ,\\
\cdots\cdots\cdots &  \, .
\nonumber
\end{align}
We shall see below that these solutions are relevant for the calculation of the rates in the toy model regime.

\subsection{Derivation of the tunneling rates}
\label{tunnelRates}
We use a rate equation formalism in order to describe the transport through the setup of a quantum dot that is coupled to a QPC. The input for the rate equation is tunneling rates. In this Appendix, we provide a full derivation of these rates up to fourth order in the tunneling elements of the dot. We work in the interaction picture with respect to $H_T$ [see Eq.~\eqref{eq: Ht}].
As we have seen in the main text, since the QPC dynamics interplays with the charge on the dot [cf. Eqs.~\eqref{emptyWF} and \eqref{fullWF}], its time evolution is also affected by $H_T$.

\emph{To zeroth order} in $H_T$, there are no tunneling events in the dot and the only dynamics are of the QPC. For example, starting from the initial state with $0$ electrons that have tunneled through the QPC and empty dot, $\ket{0\, ;e\, ; t=0}$, the probability amplitude to end up at time $t$ with $i$ electrons that have tunneled through the QPC with momenta $\mathbf{l}_i,\mathbf{r}_i$ is
\begin{align}
\braket{\mathbf{l}_i,\mathbf{r}_i \, ; \alpha \, ; t}{0\, ; e \, ; t=0}^{(0)}=b_{\mathbf{l}_i,\mathbf{r}_i}(t)\delta_{e\alpha},
\end{align}
where $\alpha$ is the charge on dot. Similarly,
\begin{align}
\braket{\mathbf{l}_i,\mathbf{r}_i \, ; \alpha \, ;  t}{0 \, ; f \, ; t=0}=\tilde{b}_{\mathbf{l}_i,\mathbf{r}_i}(t)\delta_{f\alpha}.
\end{align}

\emph{First order}. To the lowest order in $H_T$, the transition rates of the dot can be calculated using Fermi's ``golden rule''. The probability amplitude for a sequential tunneling of an electron with energy, $\epsilon$, from the source to the dot accompanied by tunneling through the QPC is
\begin{widetext}
\begin{align}
\braket{ \mathbf{l}_i,\mathbf{r}_i \, ; f \, ; t}{0\, ; e \, ; t=0}^{(1)}&=-\frac{i}{\hbar}\int_0^t \, dt' \bra{ \mathbf{l}_i,\mathbf{r}_i \, ; f } e^{-i\frac{H_0}{\hbar}t}e^{i\frac{H_0}{\hbar}t'}H_T e^{-i\frac{H_0}{\hbar}t'} \ket{0\, ; e }\nonumber\\
&=-\frac{i }{\hbar}t_S^{*}\int_0^t \, dt' \sum_{j}\int_{-\infty}^{\mu_L}d\mathbf{l}_j \rho_L^j\int_{\mu_R}^{\infty}d\mathbf{r}_j \rho_R^j e^{i\frac{(\epsilon_d-\epsilon)}{\hbar}t'}\tilde b_{\mathbf{l}_i,\mathbf{r}_i|\mathbf{l}_j,\mathbf{r}_j}(t-t')b_{\mathbf{l}_j,\mathbf{r}_j}(t'),
\end{align}
\end{widetext}
where $t_S, t_D$ are the tunneling coefficients, $b_{\mathbf{l}_j,\mathbf{r}_j}(t')$ is the probability amplitude that, up to time $t'$, $j$ electrons with momenta $\mathbf{l}_j$ were transmitted to momenta $\mathbf{r}_j$, and $b_{\mathbf{l}_i,\mathbf{r}_i|\mathbf{l}_j,\mathbf{r}_j}(t-t')$ is the time-dependent conditional probability amplitude to find the system at time $t$ in a state with $i$ electrons transmitted from left momenta $\mathbf{l}_i$ to right momenta $\mathbf{r}_i$, given that the aforementioned $j$ electrons were transmitted up to time $t'$. Similarly, tunneling from the dot to the lead reads
\begin{widetext}
\begin{align}
\braket{ \mathbf{l}_i,\mathbf{r}_i \, ; e \, ; t}{0\, ; f \, ; t=0}^{(1)}&=-\frac{i }{\hbar}\int_0^t \, dt' \bra{ \mathbf{l}_i,\mathbf{r}_i \, ; e } e^{-i\frac{H_0}{\hbar}t}e^{i\frac{H_0}{\hbar}t'}H_T e^{-i\frac{H_0}{\hbar}t'} \ket{0\, ; f }\nonumber\\
&=-\frac{i }{\hbar}t_D\int_0^t \, dt' \sum_{j}\int_{-\infty}^{\mu_L}d\mathbf{l}_j \rho_L^j\int_{\mu_R}^{\infty}d\mathbf{r}_j \rho_R^j e^{-i\frac{(\epsilon_d-\epsilon)t'}{\hbar}} b_{\mathbf{l}_i,\mathbf{r}_i|\mathbf{l}_j,\mathbf{r}_j}(t-t')\tilde b_{\mathbf{l}_j,\mathbf{r}_j}(t').
\end{align}
\end{widetext}

\emph{Second order}. To next order in $H_T$, the electron can virtually tunnel through the dot. We write the probability amplitude for a cotunneling event where an electron with energy, $\epsilon$, tunnels from the source to the dot and an electron tunnels from the dot to the drain with energy $\epsilon'$. During this process $i$ electrons tunnel through the QPC
\begin{widetext}
\begin{align}
\braket{ \mathbf{l}_i,\mathbf{r}_i \, ; e \, ; t}{0\, ; e \, ; t=0}^{(2)}=&\left(-\frac{i }{\hbar}\right)^2\int_0^t  dt'\int_0^{t'} \, dt'' \bra{ \mathbf{l}_i,\mathbf{r}_i \, ; e } e^{-i\frac{H_0}{\hbar}(t-t')}H_T e^{-i\frac{H_0}{\hbar}(t'-t'')}H_T e^{-i\frac{H_0}{\hbar}t''} \ket{0\, ; e }\nonumber\\
=&-\frac{t_S^* t_D }{\hbar^2}\int_0^t \, dt' \int_0^{t'} \, dt'' \sum_{j,k}\int_{-\infty}^{\mu_L}d\mathbf{l}_k \rho_L^k \int_{\mu_R}^{\infty}d\mathbf{r}_k \rho_R^k \int_{-\infty}^{\mu_L}d\mathbf{l}_j \rho_L^j\int_{\mu_R}^{\infty}d\mathbf{r}_j \rho_R^j\times\nonumber\\
& e^{-i\frac{(\epsilon_d-\epsilon')t'}{\hbar}} e^{-i\frac{(\epsilon-\epsilon_d)t''}{\hbar}} b_{\mathbf{l}_i,\mathbf{r}_i|\mathbf{l}_k,\mathbf{r}_k}(t-t')\tilde b_{\mathbf{l}_k,\mathbf{r}_k|\mathbf{l}_j,\mathbf{r}_j}(t'-t'') b_{\mathbf{l}_j,\mathbf{r}_j}(t''),
\end{align}
\end{widetext}
$b_{\mathbf{l}_j,\mathbf{r}_j}(t'')$ is the probability amplitude that, up to time $t''$, $j$ electrons with momenta $\mathbf{l}_j$ were transmitted to momenta $\mathbf{r}_j$ while the dot was empty, $\tilde{b}_{\mathbf{l}_k,\mathbf{r}_k|\mathbf{l}_j,\mathbf{r}_j}(t'-t'')$ is the time-dependent conditional probability amplitude to find the system at time $t'$ in a state with $k$ electrons transmitted from left momenta $\mathbf{l}_k$ to right momenta $\mathbf{r}_k$ after a time evolution from time $t''$ at a presence of a full dot, given that the aforementioned $j$ electrons were transmitted up to time $t''$, and  $b_{\mathbf{l}_i,\mathbf{r}_i|\mathbf{l}_k,\mathbf{r}_k}(t-t')$ is the time-dependent conditional probability amplitude to find the system at time $t$ in a state with $i$ electrons transmitted from left momenta $\mathbf{l}_i$ to right momenta $\mathbf{r}_i$ after a time evolution from time $t'$ at a presence of an empty dot, given that the aforementioned $k$ electrons were transmitted up to time $t'$.

Similarly, the cotunneling event of an electron with energy $\epsilon$ to tunnel in and $\epsilon'$ tunnels out with an initially full dot is
\begin{widetext}
\begin{align}
\braket{ \mathbf{l}_i,\mathbf{r}_i \, ; f \, ; t}{0\, ; f \, ; t=0}^{(2)}=&\left(-\frac{i }{\hbar}\right)^2\int_0^t  dt'\int_0^{t'} \, dt'' \bra{ \mathbf{l}_i,\mathbf{r}_i \, ; f } e^{-i\frac{H_0}{\hbar}(t-t')}H_T e^{-i\frac{H_0}{\hbar}(t'-t'')}H_T e^{-i\frac{H_0}{\hbar}t''} \ket{0\, ; f }\nonumber\\
=&-\frac{t_S^* t_D }{\hbar^2}\int_0^t \, dt' \int_0^{t'} \, dt'' \sum_{j,k}\int_{-\infty}^{\mu_L}d\mathbf{l}_k \rho_L^k\int_{\mu_R}^{\infty}d\mathbf{r}_k \rho_R^k\int_{-\infty}^{\mu_L}d\mathbf{l}_j \rho_L^j\int_{\mu_R}^{\infty}d\mathbf{r}_j \rho_R^j\times\nonumber\\
& e^{-i\frac{(\epsilon-\epsilon_d)t'}{\hbar}} e^{-i\frac{(\epsilon_d-\epsilon')t''}{\hbar}}  \tilde b_{\mathbf{l}_i,\mathbf{r}_i|\mathbf{l}_k,\mathbf{r}_k}(t-t') b_{\mathbf{l}_k,\mathbf{r}_k|\mathbf{l}_j,\mathbf{r}_j}(t'-t'') \tilde  b_{\mathbf{l}_j,\mathbf{r}_j}(t'')\,.
\label{cotFulDot}
\end{align}
\end{widetext}

Let us evaluate Eq.~\eqref{cotFulDot} for the case of the toy model described in the main text, i.e. for $\tilde{\Omega}=0$. Notice that, in this toy model regime, no transport occurs in the QPC when the dot is full. As a result, in Eq.~\eqref{cotFulDot}, we can set $j=0$ and $i=k$, leading to a cancellation of the internal energy integrals in the probability amplitude, which sum over higher-order coherent transport through the QPC.

We begin with considering the probability amplitude for a cotunneling through the dot in the presence of zero electrons passing through the QPC. 
Using Eq.~\eqref{rig10}, Eq.~\eqref{cotFulDot} in this case becomes
\begin{align}
&\braket{ 0 \, ; f \, ; t}{0\, ; f \, ; t=0}^{(2)}= \\
& -\frac{t_S^* t_D }{\hbar^2}\int_0^t \, dt' \int_0^{t'} \, dt''e^{-i\frac{(\epsilon-\epsilon_d)t'}{\hbar}} e^{-i\frac{(\epsilon_d-\epsilon')t''}{\hbar}} e^{-\frac{\mathcal{D}}{2}t}\,.\nonumber
\end{align}
Performing the time integrals, and taking the modulo square of the result yields a probability for this process to occur for equal source and drain energies $\epsilon$ (different $\epsilon$ and $\epsilon'$ have zero probability). Taking the time derivative of this probability results in the integrand of the rate appearing in Eq.~\eqref{eq: Wsd0}. 

We, now, consider the probability amplitude for a cotunneling through the dot in the presence of one electron passing through the QPC. 
Using Eqs.~\eqref{rig10}, \eqref{rig11}, and \eqref{rig13}, the expression becomes
\begin{align}
&\braket{ 0 \, ; f \, ; t}{0\, ; f \, ; t=0}^{(2)}= -\frac{t_S^* t_D }{\hbar^2}\int_0^t \, dt' \int_0^{t'} \, dt''\nonumber\\
& e^{-i\frac{(\epsilon-\epsilon_d)t'}{\hbar}} e^{-i\frac{(\epsilon_d-\epsilon')t''}{\hbar}} e^{i\frac{\left(\mathcal{E}_{l_1}-\mathcal{E}_{r_1}\right)(t-t')}{\hbar}}\times\label{appWsd1}\\
&\frac{\Omega }{\mathcal{E}_{l_1}-\mathcal{E}_{r_1}}e^{-\frac{\mathcal{D}}{2}(t'-t'')}\left[1-e^{i\frac{(\mathcal{E}_{l_1}-\mathcal{E}_{r_1})(t'-t'')}{\hbar}}\right]\,.\nonumber
\end{align}
Here, too, the evaluation of Eq.~\eqref{appWsd1} followed by modulo square of the result yields a probability for the process to occur. The obtain rate is, then, integrated upon in Eq.~\eqref{integralForm}.


\begin{thebibliography}{44}%
\makeatletter
\providecommand \@ifxundefined [1]{%
 \@ifx{#1\undefined}
}%
\providecommand \@ifnum [1]{%
 \ifnum #1\expandafter \@firstoftwo
 \else \expandafter \@secondoftwo
 \fi
}%
\providecommand \@ifx [1]{%
 \ifx #1\expandafter \@firstoftwo
 \else \expandafter \@secondoftwo
 \fi
}%
\providecommand \natexlab [1]{#1}%
\providecommand \enquote  [1]{``#1''}%
\providecommand \bibnamefont  [1]{#1}%
\providecommand \bibfnamefont [1]{#1}%
\providecommand \citenamefont [1]{#1}%
\providecommand \href@noop [0]{\@secondoftwo}%
\providecommand \href [0]{\begingroup \@sanitize@url \@href}%
\providecommand \@href[1]{\@@startlink{#1}\@@href}%
\providecommand \@@href[1]{\endgroup#1\@@endlink}%
\providecommand \@sanitize@url [0]{\catcode `\\12\catcode `\$12\catcode
  `\&12\catcode `\#12\catcode `\^12\catcode `\_12\catcode `\%12\relax}%
\providecommand \@@startlink[1]{}%
\providecommand \@@endlink[0]{}%
\providecommand \url  [0]{\begingroup\@sanitize@url \@url }%
\providecommand \@url [1]{\endgroup\@href {#1}{\urlprefix }}%
\providecommand \urlprefix  [0]{URL }%
\providecommand \Eprint [0]{\href }%
\providecommand \doibase [0]{http://dx.doi.org/}%
\providecommand \selectlanguage [0]{\@gobble}%
\providecommand \bibinfo  [0]{\@secondoftwo}%
\providecommand \bibfield  [0]{\@secondoftwo}%
\providecommand \translation [1]{[#1]}%
\providecommand \BibitemOpen [0]{}%
\providecommand \bibitemStop [0]{}%
\providecommand \bibitemNoStop [0]{.\EOS\space}%
\providecommand \EOS [0]{\spacefactor3000\relax}%
\providecommand \BibitemShut  [1]{\csname bibitem#1\endcsname}%
\let\auto@bib@innerbib\@empty
\bibitem [{\citenamefont {Helstrom}(1976)}]{Helstrom}%
  \BibitemOpen
  \bibfield  {author} {\bibinfo {author} {\bibfnamefont {C.~W.}\ \bibnamefont
  {Helstrom}},\ }\href@noop {} {\emph {\bibinfo {title} {Quantum Detection and
  Estimation Theory}}}\ (\bibinfo  {publisher} {Academic, New York},\
  \bibinfo {year} {1976})\BibitemShut {NoStop}%
\bibitem [{\citenamefont {von Neumann}(1983)}]{von-neumann}%
  \BibitemOpen
  \bibfield  {author} {\bibinfo {author} {\bibfnamefont {J.}~\bibnamefont {von
  Neumann}},\ }\href@noop {} {\emph {\bibinfo {title} {{Mathematical
  Foundations of Quantum Theory}}}}\ (\bibinfo  {publisher} {Princeton
  University Press, Princeton, NJ},\ \bibinfo {year} {1983})\BibitemShut {NoStop}%
\bibitem [{\citenamefont {Condon}\ and\ \citenamefont
  {Morse}(1931)}]{Condon1931}%
  \BibitemOpen
  \bibfield  {author} {\bibinfo {author} {\bibfnamefont {E.}~\bibnamefont
  {Condon}}\ and\ \bibinfo {author} {\bibfnamefont {P.}~\bibnamefont {Morse}},\
  }\href {\doibase 10.1103/RevModPhys.3.43} {\bibfield  {journal} {\bibinfo
  {journal} {Rev. Mod, Phys.}\ }\textbf {\bibinfo {volume} {3}},\ \bibinfo
  {pages} {43} (\bibinfo {year} {1931})}\BibitemShut {NoStop}%
\bibitem [{\citenamefont {Wigner}(1955)}]{Wigner1955}%
  \BibitemOpen
  \bibfield  {author} {\bibinfo {author} {\bibfnamefont {E.}~\bibnamefont
  {Wigner}},\ }\href {\doibase 10.1103/PhysRev.98.145} {\bibfield  {journal}
  {\bibinfo  {journal} {Phys. Rev.}\ }\textbf {\bibinfo {volume} {98}},\
  \bibinfo {pages} {145} (\bibinfo {year} {1955})}\BibitemShut {NoStop}%
\bibitem [{\citenamefont {B\"{u}ttiker}\ and\ \citenamefont
  {Landauer}(1982)}]{Buttiker1982}%
  \BibitemOpen
  \bibfield  {author} {\bibinfo {author} {\bibfnamefont {M.}~\bibnamefont
  {B\"{u}ttiker}}\ and\ \bibinfo {author} {\bibfnamefont {R.}~\bibnamefont
  {Landauer}},\ }\href {\doibase 10.1103/PhysRevLett.49.1739} {\bibfield
  {journal} {\bibinfo  {journal} {Phys. Rev. Lett.}\ }\textbf {\bibinfo
  {volume} {49}},\ \bibinfo {pages} {1739} (\bibinfo {year}
  {1982})}\BibitemShut {NoStop}%
\bibitem [{\citenamefont {Sokolovski}(1987)}]{Sokolovski1987}%
  \BibitemOpen
  \bibfield  {author} {\bibinfo {author} {\bibfnamefont {D.}~\bibnamefont
  {Sokolovski}}\ and\ \bibinfo {author} {\bibfnamefont {L.~M.}~\bibnamefont
  {Baskin}},\ }\href {\doibase 10.1103/PhysRevA.36.4604} {\bibfield
  {journal} {\bibinfo  {journal} {Phys. Rev. A}\ }\textbf {\bibinfo
  {volume} {36}},\ \bibinfo {pages} {4604} (\bibinfo {year}
  {1987})}\BibitemShut {NoStop}%
\bibitem [{\citenamefont {Steinberg}(1995)}]{Steinberg1995}%
  \BibitemOpen
  \bibfield  {author} {\bibinfo {author} {\bibfnamefont {A.~M.}~\bibnamefont
  {Steinberg}},\ }\href {\doibase 10.1103/PhysRevLett.74.2405} {\bibfield
  {journal} {\bibinfo  {journal} {Phys. Rev. Lett.}\ }\textbf {\bibinfo
  {volume} {74}},\ \bibinfo {pages} {2405} (\bibinfo {year}
  {1995})}\BibitemShut {NoStop}%
\bibitem [{\citenamefont {Clerk}\ \emph {et~al.}(2010)\citenamefont {Clerk},
  \citenamefont {Girvin}, \citenamefont {Marquardt},\ and\ \citenamefont
  {Schoelkopf}}]{Clerk2010}%
  \BibitemOpen
  \bibfield  {author} {\bibinfo {author} {\bibfnamefont {A.~A.}\ \bibnamefont
  {Clerk}}, \bibinfo {author} {\bibfnamefont {S.~M.}\ \bibnamefont {Girvin}},
  \bibinfo {author} {\bibfnamefont {F.}~\bibnamefont {Marquardt}}, \ and\
  \bibinfo {author} {\bibfnamefont {R.~J.}\ \bibnamefont {Schoelkopf}},\ }\href
  {\doibase 10.1103/RevModPhys.82.1155} {\bibfield  {journal} {\bibinfo
  {journal} {Rev. Mod. Phys.}\ }\textbf {\bibinfo {volume} {82}},\ \bibinfo
  {pages} {1155} (\bibinfo {year} {2010})}\BibitemShut {NoStop}%
\bibitem [{\citenamefont {Aharonov}\ \emph
  {et~al.}(1988{\natexlab{a}})\citenamefont {Aharonov}, \citenamefont
  {Albert},\ and\ \citenamefont {Vaidman}}]{Aharonov:1988a}%
  \BibitemOpen
  \bibfield  {author} {\bibinfo {author} {\bibfnamefont {Y.}~\bibnamefont
  {Aharonov}}, \bibinfo {author} {\bibfnamefont {D.~Z.}~\bibnamefont {Albert}}, \
  and\ \bibinfo {author} {\bibfnamefont {L.}~\bibnamefont {Vaidman}},\
  }\href@noop {} {\bibfield  {journal} {\bibinfo  {journal} {Phys. Rev. Lett.}\
  }\textbf {\bibinfo {volume} {60}},\ \bibinfo {pages} {1351} (\bibinfo {year}
  {1988}{\natexlab{a}})}\BibitemShut {NoStop}%
\bibitem [{\citenamefont {Korotkov}(2001)}]{Korotkov:2001a}%
  \BibitemOpen
  \bibfield  {author} {\bibinfo {author} {\bibfnamefont {A.~N.}\ \bibnamefont
  {Korotkov}},\ }\href@noop {} {\bibfield  {journal} {\bibinfo  {journal}
  {Phys. Rev. B}\ }\textbf {\bibinfo {volume} {63}},\ \bibinfo {pages} {115403}
  (\bibinfo {year} {2001})}\BibitemShut {NoStop}%
\bibitem [{\citenamefont {Korotkov}\ and\ \citenamefont
  {Averin}(2001)}]{Korotkov:2001b}%
  \BibitemOpen
  \bibfield  {author} {\bibinfo {author} {\bibfnamefont {A.~N.}\ \bibnamefont
  {Korotkov}}\ and\ \bibinfo {author} {\bibfnamefont {D.~V.}\ \bibnamefont
  {Averin}},\ }\href@noop {} {\bibfield  {journal} {\bibinfo  {journal} {Phys.
  Rev. B}\ }\textbf {\bibinfo {volume} {64}},\ \bibinfo {pages} {165310}
  (\bibinfo {year} {2001})}\BibitemShut {NoStop}%
\bibitem [{\citenamefont {Hosten}\ and\ \citenamefont
  {Kwiat}(2008)}]{Hosten:2008}%
  \BibitemOpen
  \bibfield  {author} {\bibinfo {author} {\bibfnamefont {O.}~\bibnamefont
  {Hosten}}\ and\ \bibinfo {author} {\bibfnamefont {P.}~\bibnamefont {Kwiat}},\
  }\href@noop {} {\bibfield  {journal} {\bibinfo  {journal} {Science}\ }\textbf
  {\bibinfo {volume} {319}},\ \bibinfo {pages} {787} (\bibinfo {year}
  {2008})}\BibitemShut {NoStop}%
\bibitem [{\citenamefont {Dixon}\ \emph {et~al.}(2009)\citenamefont {Dixon},
  \citenamefont {Starling}, \citenamefont {Jordan},\ and\ \citenamefont
  {Howell}}]{Dixon:2009}%
  \BibitemOpen
  \bibfield  {author} {\bibinfo {author} {\bibfnamefont {P.~B.}\ \bibnamefont
  {Dixon}}, \bibinfo {author} {\bibfnamefont {D.~J.}\ \bibnamefont {Starling}},
  \bibinfo {author} {\bibfnamefont {A.~N.}\ \bibnamefont {Jordan}}, \ and\
  \bibinfo {author} {\bibfnamefont {J.~C.}\ \bibnamefont {Howell}},\
  }\href@noop {} {\bibfield  {journal} {\bibinfo  {journal} {Phys. Rev. Lett.}\
  }\textbf {\bibinfo {volume} {102}},\ \bibinfo {pages} {173601} (\bibinfo
  {year} {2009})}\BibitemShut {NoStop}%
\bibitem [{\citenamefont {Starling}\ \emph {et~al.}(2009)\citenamefont
  {Starling}, \citenamefont {Dixon}, \citenamefont {Jordan},\ and\
  \citenamefont {Howell}}]{Starling:2009}%
  \BibitemOpen
  \bibfield  {author} {\bibinfo {author} {\bibfnamefont {D.~J.}~\bibnamefont
  {Starling}}, \bibinfo {author} {\bibfnamefont {P.~B.}~\bibnamefont {Dixon}},
  \bibinfo {author} {\bibfnamefont {A.~N.}~\bibnamefont {Jordan}}, \ and\ \bibinfo
  {author} {\bibfnamefont {J.~C.}~\bibnamefont {Howell}},\ }\href@noop {}
  {\bibfield  {journal} {\bibinfo  {journal} {Phys. Rev. A}\ }\textbf {\bibinfo
  {volume} {80}},\ \bibinfo {pages} {041803} (\bibinfo {year}
  {2009})}\BibitemShut {NoStop}%
\bibitem [{\citenamefont {Brunner}\ and\ \citenamefont
  {Simon}(2010)}]{Brunner:2010}%
  \BibitemOpen
  \bibfield  {author} {\bibinfo {author} {\bibfnamefont {N.}~\bibnamefont
  {Brunner}}\ and\ \bibinfo {author} {\bibfnamefont {C.}~\bibnamefont
  {Simon}},\ }\href@noop {} {\bibfield  {journal} {\bibinfo  {journal} {Phys.
  Rev. Lett.}\ }\textbf {\bibinfo {volume} {105}},\ \bibinfo {pages} {010405}
  (\bibinfo {year} {2010})}\BibitemShut {NoStop}%
\bibitem [{\citenamefont {Starling}\ \emph {et~al.}(2010)\citenamefont
  {Starling}, \citenamefont {Dixon}, \citenamefont {Williams}, \citenamefont
  {Jordan},\ and\ \citenamefont {Howell}}]{Starling:2010b}%
  \BibitemOpen
  \bibfield  {author} {\bibinfo {author} {\bibfnamefont {D.~J.}\ \bibnamefont
  {Starling}}, \bibinfo {author} {\bibfnamefont {P.~B.}\ \bibnamefont {Dixon}},
  \bibinfo {author} {\bibfnamefont {N.~S.}\ \bibnamefont {Williams}}, \bibinfo
  {author} {\bibfnamefont {A.~N.}\ \bibnamefont {Jordan}}, \ and\ \bibinfo
  {author} {\bibfnamefont {J.~C.}\ \bibnamefont {Howell}},\ }\href@noop {}
  {\bibfield  {journal} {\bibinfo  {journal} {Phys. Rev. A}\ }\textbf {\bibinfo
  {volume} {82}},\ \bibinfo {pages} {011802} (\bibinfo {year}
  {2010})}\BibitemShut {NoStop}%
\bibitem [{\citenamefont {Zilberberg}\ \emph {et~al.}(2011)\citenamefont
  {Zilberberg}, \citenamefont {Romito},\ and\ \citenamefont
  {Gefen}}]{Zilberberg:2011}%
  \BibitemOpen
  \bibfield  {author} {\bibinfo {author} {\bibfnamefont {O.}~\bibnamefont
  {Zilberberg}}, \bibinfo {author} {\bibfnamefont {A.}~\bibnamefont {Romito}},
  \ and\ \bibinfo {author} {\bibfnamefont {Y.}~\bibnamefont {Gefen}},\
  }\href@noop {} {\bibfield  {journal} {\bibinfo  {journal} {Phys. Rev. Lett.}\
  }\textbf {\bibinfo {volume} {106}},\ \bibinfo {pages} {080405} (\bibinfo
  {year} {2011})}\BibitemShut {NoStop}%
\bibitem [{\citenamefont {Zilberberg}\ \emph {et~al.}(2013)\citenamefont
  {Zilberberg}, \citenamefont {Romito}, \citenamefont {Starling}, \citenamefont
  {Howland}, \citenamefont {Broadbent}, \citenamefont {Howell},\ and\
  \citenamefont {Gefen}}]{Zilberberg2013}%
  \BibitemOpen
  \bibfield  {author} {\bibinfo {author} {\bibfnamefont {O.}~\bibnamefont
  {Zilberberg}}, \bibinfo {author} {\bibfnamefont {A.}~\bibnamefont {Romito}},
  \bibinfo {author} {\bibfnamefont {D.~J.}\ \bibnamefont {Starling}}, \bibinfo
  {author} {\bibfnamefont {G.~A.}\ \bibnamefont {Howland}}, \bibinfo {author}
  {\bibfnamefont {C.~J.}\ \bibnamefont {Broadbent}}, \bibinfo {author}
  {\bibfnamefont {J.~C.}\ \bibnamefont {Howell}}, \ and\ \bibinfo {author}
  {\bibfnamefont {Y.}~\bibnamefont {Gefen}},\ }\href {\doibase
  10.1103/PhysRevLett.110.170405} {\bibfield  {journal} {\bibinfo  {journal}
  {Phys. Rev. Lett.}\ }\textbf {\bibinfo {volume} {110}},\ \bibinfo {pages}
  {170405} (\bibinfo {year} {2013})}\BibitemShut {NoStop}%
\bibitem [{\citenamefont {B\"{u}ttiker}(1983)}]{Buttiker1983}%
  \BibitemOpen
  \bibfield  {author} {\bibinfo {author} {\bibfnamefont {M.}~\bibnamefont
  {B\"{u}ttiker}},\ }\href {\doibase 10.1103/PhysRevB.27.6178} {\bibfield
  {journal} {\bibinfo  {journal} {Phys. Rev. B}\ }\textbf {\bibinfo
  {volume} {27}},\ \bibinfo {pages} {6178} (\bibinfo {year}
  {1983})}\BibitemShut {NoStop}%
\bibitem [{\citenamefont {Romito}\ and\ \citenamefont
  {Gefen}(2014)}]{Romito2014}%
  \BibitemOpen
  \bibfield  {author} {\bibinfo {author} {\bibfnamefont {A.}~\bibnamefont
  {Romito}}\ and\ \bibinfo {author} {\bibfnamefont {Y.}~\bibnamefont {Gefen}},\
  }\href {\doibase 10.1103/PhysRevB.90.085417} {\bibfield  {journal} {\bibinfo
  {journal} {Phys. Rev. B}\ }\textbf {\bibinfo {volume} {90}},\ \bibinfo
  {pages} {085417} (\bibinfo {year} {2014})}\BibitemShut {NoStop}%
\bibitem [{\citenamefont {Aleiner}\ \emph {et~al.}(2002)\citenamefont
  {Aleiner}, \citenamefont {Brouwer},\ and\ \citenamefont
  {Glazman}}]{Aleiner2002}%
  \BibitemOpen
  \bibfield  {author} {\bibinfo {author} {\bibfnamefont {I.~L.}\ \bibnamefont
  {Aleiner}}, \bibinfo {author} {\bibfnamefont {P.~W.}\ \bibnamefont
  {Brouwer}}, \ and\ \bibinfo {author} {\bibfnamefont {L.~I.}\ \bibnamefont
  {Glazman}},\ }\href {PhysRep.358.309} {\bibfield  {journal} {\bibinfo
  {journal} {Phys. Rep.}\ }\textbf {\bibinfo {volume} {358}},\ \bibinfo
  {pages} {309} (\bibinfo {year} {2002})}\BibitemShut {NoStop}%
\bibitem [{\citenamefont {Glazman}\ and\ \citenamefont
  {Pustilnik}(2005)}]{Glazman2005}%
  \BibitemOpen
  \bibfield  {author} {\bibinfo {author} {\bibfnamefont {L.~I.}~\bibnamefont
  {Glazman}}\ and\ \bibinfo {author} {\bibfnamefont {M.}~\bibnamefont
  {Pustilnik}},\ }in\ \href@noop {} {\emph {\bibinfo {booktitle} {Nanophysics:
  Coherence and transport}}},\ \bibinfo {series and number} {Lecture Notes of
  the Les Houches Summer School 2004}\ (\bibinfo  {publisher} {Elsevier, Amsterdam},\
  \bibinfo {year} {2005})\BibitemShut
  {NoStop}%
\bibitem [{\citenamefont {Field}\ \emph {et~al.}(1993)\citenamefont {Field},
  \citenamefont {Smith}, \citenamefont {Pepper}, \citenamefont {Ritchie},
  \citenamefont {Frost}, \citenamefont {Jones},\ and\ \citenamefont
  {Hasko}}]{Field1993}%
  \BibitemOpen
  \bibfield  {author} {\bibinfo {author} {\bibfnamefont {M.}~\bibnamefont
  {Field}}, \bibinfo {author} {\bibfnamefont {C.~G.}~\bibnamefont {Smith}},
  \bibinfo {author} {\bibfnamefont {M.}~\bibnamefont {Pepper}}, \bibinfo
  {author} {\bibfnamefont {D.~A.}~\bibnamefont {Ritchie}}, \bibinfo {author}
  {\bibfnamefont {J.~E.~F.}~\bibnamefont {Frost}}, \bibinfo {author} {\bibfnamefont
  {G.~A.~C.}~\bibnamefont {Jones}}, \ and\ \bibinfo {author} {\bibfnamefont
  {D.~G.}~\bibnamefont {Hasko}},\ }\href {\doibase 10.1103/PhysRevLett.70.1311}
  {\bibfield  {journal} {\bibinfo  {journal} {Phys. Rev. Lett.}\ }\textbf
  {\bibinfo {volume} {70}},\ \bibinfo {pages} {1311} (\bibinfo {year}
  {1993})}\BibitemShut {NoStop}%
\bibitem [{\citenamefont {Elzerman}\ \emph {et~al.}(2003)\citenamefont
  {Elzerman}, \citenamefont {Hanson}, \citenamefont {Greidanus}, \citenamefont
  {Willems~van Beveren}, \citenamefont {De~Franceschi}, \citenamefont
  {Vandersypen}, \citenamefont {Tarucha},\ and\ \citenamefont
  {Kouwenhoven}}]{Elzerman2003}%
  \BibitemOpen
  \bibfield  {author} {\bibinfo {author} {\bibfnamefont {J.~M.}\ \bibnamefont
  {Elzerman}}, \bibinfo {author} {\bibfnamefont {R.}~\bibnamefont {Hanson}},
  \bibinfo {author} {\bibfnamefont {J.~S.}\ \bibnamefont {Greidanus}}, \bibinfo
  {author} {\bibfnamefont {L.~H.~Willems}\ \bibnamefont {van Beveren}},
  \bibinfo {author} {\bibfnamefont {S.}~\bibnamefont {De~Franceschi}}, \bibinfo
  {author} {\bibfnamefont {L.~M.~K.}\ \bibnamefont {Vandersypen}}, \bibinfo
  {author} {\bibfnamefont {S.}~\bibnamefont {Tarucha}}, \ and\ \bibinfo
  {author} {\bibfnamefont {L.~P.}\ \bibnamefont {Kouwenhoven}},\ }\href@noop {}
  {\bibfield  {journal} {\bibinfo  {journal} {Phys. Rev. B}\ }\textbf {\bibinfo
  {volume} {67}},\ \bibinfo {pages} {161308} (\bibinfo {year}
  {2003})}\BibitemShut {NoStop}%
\bibitem [{\citenamefont {DiCarlo}\ \emph {et~al.}(2004)\citenamefont
  {DiCarlo}, \citenamefont {Lynch}, \citenamefont {Johnson}, \citenamefont
  {Childress}, \citenamefont {Crockett}, \citenamefont {Marcus}, \citenamefont
  {Hanson},\ and\ \citenamefont {Gossard}}]{DiCarlo2004}%
  \BibitemOpen
  \bibfield  {author} {\bibinfo {author} {\bibfnamefont {L.}~\bibnamefont
  {DiCarlo}}, \bibinfo {author} {\bibfnamefont {H.~J.}~\bibnamefont {Lynch}},
  \bibinfo {author} {\bibfnamefont {A.~C.}~\bibnamefont {Johnson}}, \bibinfo
  {author} {\bibfnamefont {L.~I.}~\bibnamefont {Childress}}, \bibinfo {author}
  {\bibfnamefont {K.}~\bibnamefont {Crockett}}, \bibinfo {author}
  {\bibfnamefont {C.~M.}~\bibnamefont {Marcus}}, \bibinfo {author} {\bibfnamefont
  {M.~P.}~\bibnamefont {Hanson}}, \ and\ \bibinfo {author} {\bibfnamefont
  {A.~C.}~\bibnamefont {Gossard}},\ }\href {\doibase
  10.1103/PhysRevLett.92.226801} {\bibfield  {journal} {\bibinfo  {journal}
  {Phys. Rev. Lett.}\ }\textbf {\bibinfo {volume} {92}},\ \bibinfo {pages}
  {226801} (\bibinfo {year} {2004})}\BibitemShut {NoStop}%
\bibitem [{\citenamefont {Harbusch}\ \emph {et~al.}(2010)\citenamefont
  {Harbusch}, \citenamefont {Taubert}, \citenamefont {Tranitz}, \citenamefont
  {Wegscheider},\ and\ \citenamefont {Ludwig}}]{Harbusch2010}%
  \BibitemOpen
  \bibfield  {author} {\bibinfo {author} {\bibfnamefont {D.}~\bibnamefont
  {Harbusch}}, \bibinfo {author} {\bibfnamefont {D.}~\bibnamefont {Taubert}},
  \bibinfo {author} {\bibfnamefont {H.~P.}\ \bibnamefont {Tranitz}}, \bibinfo
  {author} {\bibfnamefont {W.}~\bibnamefont {Wegscheider}}, \ and\ \bibinfo
  {author} {\bibfnamefont {S.}~\bibnamefont {Ludwig}},\ }\href@noop {}
  {\bibfield  {journal} {\bibinfo  {journal} {Phys. Rev. Lett.}\ }\textbf
  {\bibinfo {volume} {104}},\ \bibinfo {pages} {196801} (\bibinfo {year}
  {2010})}\BibitemShut {NoStop}%
\bibitem [{\citenamefont {Gasparinetti}\ \emph {et~al.}(2012)\citenamefont
  {Gasparinetti}, \citenamefont {Martínez-Pérez}, \citenamefont
  {de~Franceschi}, \citenamefont {Pekola},\ and\ \citenamefont
  {Giazotto}}]{Gasparinetti2012}%
  \BibitemOpen
  \bibfield  {author} {\bibinfo {author} {\bibfnamefont {S.}~\bibnamefont
  {Gasparinetti}}, \bibinfo {author} {\bibfnamefont {M.~J.}\ \bibnamefont
  {Martínez-Pérez}}, \bibinfo {author} {\bibfnamefont {S.}~\bibnamefont
  {de~Franceschi}}, \bibinfo {author} {\bibfnamefont {J.~P.}\ \bibnamefont
  {Pekola}}, \ and\ \bibinfo {author} {\bibfnamefont {F.}~\bibnamefont
  {Giazotto}},\ }\href@noop {} {\bibfield  {journal} {\bibinfo  {journal}
  {Appl. Phys. Lett.}\ }\textbf {\bibinfo {volume} {100}},\ \bibinfo {pages}
  {253502} (\bibinfo {year} {2012})}\BibitemShut {NoStop}%
\bibitem [{\citenamefont {Granger}\ \emph {et~al.}(2012)\citenamefont
  {Granger}, \citenamefont {Taubert}, \citenamefont {Young}, \citenamefont
  {Gaudreau}, \citenamefont {Kam}, \citenamefont {Studenikin}, \citenamefont
  {Zawadzki}, \citenamefont {Harbusch}, \citenamefont {Schuh}, \citenamefont
  {Wegscheider}, \citenamefont {Wasilewski}, \citenamefont {Clerk},
  \citenamefont {Ludwig},\ and\ \citenamefont {Sachrajda}}]{Granger2012}%
  \BibitemOpen
  \bibfield  {author} {\bibinfo {author} {\bibfnamefont {G.}~\bibnamefont
  {Granger}}, \bibinfo {author} {\bibfnamefont {D.}~\bibnamefont {Taubert}},
  \bibinfo {author} {\bibfnamefont {C.~E.}\ \bibnamefont {Young}}, \bibinfo
  {author} {\bibfnamefont {L.}~\bibnamefont {Gaudreau}}, \bibinfo {author}
  {\bibfnamefont {A.}~\bibnamefont {Kam}}, \bibinfo {author} {\bibfnamefont
  {S.~A.}\ \bibnamefont {Studenikin}}, \bibinfo {author} {\bibfnamefont
  {P.}~\bibnamefont {Zawadzki}}, \bibinfo {author} {\bibfnamefont
  {D.}~\bibnamefont {Harbusch}}, \bibinfo {author} {\bibfnamefont
  {D.}~\bibnamefont {Schuh}}, \bibinfo {author} {\bibfnamefont
  {W.}~\bibnamefont {Wegscheider}}, \bibinfo {author} {\bibfnamefont {Z.~R.}\
  \bibnamefont {Wasilewski}}, \bibinfo {author} {\bibfnamefont {A.~A.}\
  \bibnamefont {Clerk}}, \bibinfo {author} {\bibfnamefont {S.}~\bibnamefont
  {Ludwig}}, \ and\ \bibinfo {author} {\bibfnamefont {A.~S.}\ \bibnamefont
  {Sachrajda}},\ }\href@noop {} {\bibfield  {journal} {\bibinfo  {journal}
  {Nat. Phys.}\ }\textbf {\bibinfo {volume} {8}},\ \bibinfo {pages} {522}
  (\bibinfo {year} {2012})}\BibitemShut {NoStop}%
\bibitem [{\citenamefont {K\"{u}ng}\ \emph {et~al.}(2012)\citenamefont
  {K\"{u}ng}, \citenamefont {R\"{o}ssler}, \citenamefont {Beck}, \citenamefont
  {Faist}, \citenamefont {Ihn},\ and\ \citenamefont {Ensslin}}]{clemens2012}%
  \BibitemOpen
  \bibfield  {author} {\bibinfo {author} {\bibfnamefont {B.}~\bibnamefont
  {K\"{u}ng}}, \bibinfo {author} {\bibfnamefont {C.}~\bibnamefont
  {R\"{o}ssler}}, \bibinfo {author} {\bibfnamefont {M.}~\bibnamefont {Beck}},
  \bibinfo {author} {\bibfnamefont {J.}~\bibnamefont {Faist}}, \bibinfo
  {author} {\bibfnamefont {T.}~\bibnamefont {Ihn}}, \ and\ \bibinfo {author}
  {\bibfnamefont {K.}~\bibnamefont {Ensslin}},\ }\href@noop {} {\bibfield
  {journal} {\bibinfo  {journal} {New J. of Phys.}\ }\textbf {\bibinfo
  {volume} {14}},\ \bibinfo {pages} {083003} (\bibinfo {year}
  {2012})}\BibitemShut {NoStop}%
\bibitem [{\citenamefont {R\"{o}ssler}\ \emph {et~al.}(2013)\citenamefont
  {R\"{o}ssler}, \citenamefont {Kr\"{a}henmann}, \citenamefont {Baer},
  \citenamefont {Ihn}, \citenamefont {Ensslin}, \citenamefont {Reichl},\ and\
  \citenamefont {Wegscheider}}]{clemens2013}%
  \BibitemOpen
  \bibfield  {author} {\bibinfo {author} {\bibfnamefont {C.}~\bibnamefont
  {R\"{o}ssler}}, \bibinfo {author} {\bibfnamefont {T.}~\bibnamefont
  {Kr\"{a}henmann}}, \bibinfo {author} {\bibfnamefont {S.}~\bibnamefont
  {Baer}}, \bibinfo {author} {\bibfnamefont {T.}~\bibnamefont {Ihn}}, \bibinfo
  {author} {\bibfnamefont {K.}~\bibnamefont {Ensslin}}, \bibinfo {author}
  {\bibfnamefont {C.}~\bibnamefont {Reichl}}, \ and\ \bibinfo {author}
  {\bibfnamefont {W.}~\bibnamefont {Wegscheider}},\ }\href@noop {} {\bibfield
  {journal} {\bibinfo  {journal} {New J. of Phys.}\ }\textbf {\bibinfo
  {volume} {15}},\ \bibinfo {pages} {033011} (\bibinfo {year}
  {2013})}\BibitemShut {NoStop}%
\bibitem [{\citenamefont {Maradan}\ \emph {et~al.}(2014)\citenamefont
  {Maradan}, \citenamefont {Casparis}, \citenamefont {Liu}, \citenamefont
  {Biesinger}, \citenamefont {Scheller}, \citenamefont {Zumb\"{u}hl},
  \citenamefont {Zimmerman},\ and\ \citenamefont {Gossard}}]{zumbuhl2014}%
  \BibitemOpen
  \bibfield  {author} {\bibinfo {author} {\bibfnamefont {D.}~\bibnamefont
  {Maradan}}, \bibinfo {author} {\bibfnamefont {L.}~\bibnamefont {Casparis}},
  \bibinfo {author} {\bibfnamefont {T.-M.}\ \bibnamefont {Liu}}, \bibinfo
  {author} {\bibfnamefont {D.~E.~F.}\ \bibnamefont {Biesinger}}, \bibinfo
  {author} {\bibfnamefont {C.~P.}\ \bibnamefont {Scheller}}, \bibinfo {author}
  {\bibfnamefont {D.~M.}\ \bibnamefont {Zumb\"{u}hl}}, \bibinfo {author}
  {\bibfnamefont {J.}~\bibnamefont {Zimmerman}}, \ and\ \bibinfo {author}
  {\bibfnamefont {A.~C.}\ \bibnamefont {Gossard}},\ }\href@noop {} {\bibfield
  {journal} {\bibinfo  {journal} {J. Low Temp. Phys.}\ }\textbf {\bibinfo
  {volume} {175}},\ \bibinfo {pages} {784} (\bibinfo {year}
  {2014})}\BibitemShut
  {NoStop}%
\bibitem [{\citenamefont {Gurvitz}(1997)}]{gurvitz}%
  \BibitemOpen
  \bibfield  {author} {\bibinfo {author} {\bibfnamefont {S.~A.}\ \bibnamefont
  {Gurvitz}},\ }\href@noop {} {\bibfield  {journal} {\bibinfo  {journal} {Phys.
  Rev. B}\ }\textbf {\bibinfo {volume} {56}},\ \bibinfo {pages} {15215}
  (\bibinfo {year} {1997})}\BibitemShut {NoStop}%
\bibitem [{\citenamefont {Romito}\ \emph {et~al.}(2008)\citenamefont {Romito},
  \citenamefont {Gefen},\ and\ \citenamefont {Blanter}}]{Romito:2008}%
  \BibitemOpen
  \bibfield  {author} {\bibinfo {author} {\bibfnamefont {A.}~\bibnamefont
  {Romito}}, \bibinfo {author} {\bibfnamefont {Y.}~\bibnamefont {Gefen}}, \
  and\ \bibinfo {author} {\bibfnamefont {Y.~M.}\ \bibnamefont {Blanter}},\
  }\href@noop {} {\bibfield  {journal} {\bibinfo  {journal} {Phys. Rev. Lett.}\
  }\textbf {\bibinfo {volume} {100}},\ \bibinfo {pages} {056801} (\bibinfo
  {year} {2008})}\BibitemShut {NoStop}%
\bibitem [{\citenamefont {Averin}\ and\ \citenamefont
  {Nazarov}(1990)}]{Averin1990}%
  \BibitemOpen
  \bibfield  {author} {\bibinfo {author} {\bibfnamefont {D.~V.}~\bibnamefont
  {Averin}}\ and\ \bibinfo {author} {\bibfnamefont {Y.~V.}~\bibnamefont
  {Nazarov}},\ }\href {\doibase 10.1103/PhysRevLett.65.2446} {\bibfield
  {journal} {\bibinfo  {journal} {Phys. Rev. Lett.}\ }\textbf {\bibinfo
  {volume} {65}},\ \bibinfo {pages} {2446} (\bibinfo {year}
  {1990})}\BibitemShut {NoStop}%
\bibitem [{\citenamefont {Sukhorukov}\ \emph {et~al.}(2007)\citenamefont
  {Sukhorukov}, \citenamefont {Jordan}, \citenamefont {Gustavsson},
  \citenamefont {Leturcq}, \citenamefont {Ihn},\ and\ \citenamefont
  {Ensslin}}]{Sukhorukov2007}%
  \BibitemOpen
  \bibfield  {author} {\bibinfo {author} {\bibfnamefont {E.~V.}\ \bibnamefont
  {Sukhorukov}}, \bibinfo {author} {\bibfnamefont {A.~N.}\ \bibnamefont
  {Jordan}}, \bibinfo {author} {\bibfnamefont {S.}~\bibnamefont {Gustavsson}},
  \bibinfo {author} {\bibfnamefont {R.}~\bibnamefont {Leturcq}}, \bibinfo
  {author} {\bibfnamefont {T.}~\bibnamefont {Ihn}}, \ and\ \bibinfo {author}
  {\bibfnamefont {K.}~\bibnamefont {Ensslin}},\ }\href {\doibase
  10.1038/nphys564} {\bibfield  {journal} {\bibinfo  {journal} {Nature
  Physics}\ }\textbf {\bibinfo {volume} {3}},\ \bibinfo {pages} {243} (\bibinfo
  {year} {2007})}\BibitemShut {NoStop}%
\bibitem [{\citenamefont {Sakurai}\ and\ \citenamefont
  {Tuan}(1985)}]{sakurai1985}%
  \BibitemOpen
  \bibfield  {author} {\bibinfo {author} {\bibfnamefont {J.~J.}\ \bibnamefont
  {Sakurai}}\ and\ \bibinfo {author} {\bibfnamefont {S.~F.}\ \bibnamefont
  {Tuan}},\ }\href@noop {} {\emph {\bibinfo {title} {Modern Quantum
  Mechanics}}},\ (\bibinfo  {publisher}
  {Addison-Wesley, Reading, MA},\ \bibinfo {year} {1985})\BibitemShut
  {NoStop}%
\bibitem [{\citenamefont {Korotkov}(1994)}]{Korotkov:1994}%
  \BibitemOpen
  \bibfield  {author} {\bibinfo {author} {\bibfnamefont {A.~N.}\ \bibnamefont
  {Korotkov}},\ }\href@noop {} {\bibfield  {journal} {\bibinfo  {journal}
  {Phys. Rev. B}\ }\textbf {\bibinfo {volume} {49}},\ \bibinfo {pages} {10381}
  (\bibinfo {year} {1994})}\BibitemShut {NoStop}%
\bibitem [{\citenamefont {Koch}\ \emph {et~al.}(2006)\citenamefont {Koch},
  \citenamefont {von Oppen},\ and\ \citenamefont {Andreev}}]{Koch:2006}%
  \BibitemOpen
  \bibfield  {author} {\bibinfo {author} {\bibfnamefont {J.}~\bibnamefont
  {Koch}}, \bibinfo {author} {\bibfnamefont {F.}~\bibnamefont {von Oppen}}, \
  and\ \bibinfo {author} {\bibfnamefont {A.~V.}\ \bibnamefont {Andreev}},\
  }\href@noop {} {\bibfield  {journal} {\bibinfo  {journal} {Phys. Rev. B}\
  }\textbf {\bibinfo {volume} {74}},\ \bibinfo {pages} {205438} (\bibinfo
  {year} {2006})}\BibitemShut {NoStop}%
\bibitem [{\citenamefont {Katz}\ \emph {et~al.}(2006)\citenamefont {Katz},
  \citenamefont {Ansmann}, \citenamefont {Bialczak}, \citenamefont {Lucero},
  \citenamefont {McDermott}, \citenamefont {Neeley}, \citenamefont {Steffen},
  \citenamefont {Weig}, \citenamefont {Cleland}, \citenamefont {Martinis},\
  and\ \citenamefont {Korotkov}}]{Katz2006}%
  \BibitemOpen
  \bibfield  {author} {\bibinfo {author} {\bibfnamefont {N.}~\bibnamefont
  {Katz}}, \bibinfo {author} {\bibfnamefont {M.}~\bibnamefont {Ansmann}},
  \bibinfo {author} {\bibfnamefont {R.~C.}\ \bibnamefont {Bialczak}}, \bibinfo
  {author} {\bibfnamefont {E.}~\bibnamefont {Lucero}}, \bibinfo {author}
  {\bibfnamefont {R.}~\bibnamefont {McDermott}}, \bibinfo {author}
  {\bibfnamefont {M.}~\bibnamefont {Neeley}}, \bibinfo {author} {\bibfnamefont
  {M.}~\bibnamefont {Steffen}}, \bibinfo {author} {\bibfnamefont {E.~M.}\
  \bibnamefont {Weig}}, \bibinfo {author} {\bibfnamefont {A.~N.}\ \bibnamefont
  {Cleland}}, \bibinfo {author} {\bibfnamefont {J.~M.}\ \bibnamefont
  {Martinis}}, \ and\ \bibinfo {author} {\bibfnamefont {A.~N.}\ \bibnamefont
  {Korotkov}},\ }\href {\doibase 10.1126/science.1126475} {\bibfield  {journal}
  {\bibinfo  {journal} {Science (New York)}\ }\textbf {\bibinfo {volume}
  {312}},\ \bibinfo {pages} {1498} (\bibinfo {year} {2006})}\BibitemShut
  {NoStop}%
\bibitem [{\citenamefont {Pryadko}\ and\ \citenamefont
  {Korotkov}(2007)}]{Korotkov:2007}%
  \BibitemOpen
  \bibfield  {author} {\bibinfo {author} {\bibfnamefont {L.~P.}\ \bibnamefont
  {Pryadko}}\ and\ \bibinfo {author} {\bibfnamefont {A.~N.}\ \bibnamefont
  {Korotkov}},\ }\href@noop {} {\bibfield  {journal} {\bibinfo  {journal}
  {Phys. Rev. B}\ }\textbf {\bibinfo {volume} {76}},\ \bibinfo {pages} {100503}
  (\bibinfo {year} {2007})}\BibitemShut {NoStop}%
\bibitem [{\citenamefont {Katz}\ \emph {et~al.}(2008)\citenamefont {Katz},
  \citenamefont {Neeley}, \citenamefont {Ansmann}, \citenamefont {Bialczak},
  \citenamefont {Hofheinz}, \citenamefont {Lucero}, \citenamefont {O'Connell},
  \citenamefont {Wang}, \citenamefont {Cleland}, \citenamefont {Martinis},\
  and\ \citenamefont {Korotkov}}]{Katz08}%
  \BibitemOpen
  \bibfield  {author} {\bibinfo {author} {\bibfnamefont {N.}~\bibnamefont
  {Katz}}, \bibinfo {author} {\bibfnamefont {M.}~\bibnamefont {Neeley}},
  \bibinfo {author} {\bibfnamefont {M.}~\bibnamefont {Ansmann}}, \bibinfo
  {author} {\bibfnamefont {R.~C.}\ \bibnamefont {Bialczak}}, \bibinfo {author}
  {\bibfnamefont {M.}~\bibnamefont {Hofheinz}}, \bibinfo {author}
  {\bibfnamefont {E.}~\bibnamefont {Lucero}}, \bibinfo {author} {\bibfnamefont
  {A.}~\bibnamefont {O'Connell}}, \bibinfo {author} {\bibfnamefont
  {H.}~\bibnamefont {Wang}}, \bibinfo {author} {\bibfnamefont {A.~N.}\
  \bibnamefont {Cleland}}, \bibinfo {author} {\bibfnamefont {J.~M.}\
  \bibnamefont {Martinis}}, \ and\ \bibinfo {author} {\bibfnamefont {A.~N.}\
  \bibnamefont {Korotkov}},\ }\href@noop {} {\bibfield  {journal} {\bibinfo
  {journal} {Phys. Rev. Lett.}\ }\textbf {\bibinfo {volume} {101}},\ \bibinfo
  {pages} {200401} (\bibinfo {year} {2008})}\BibitemShut {NoStop}%
\bibitem [{\citenamefont {Zilberberg}\ \emph {et~al.}(2012)\citenamefont
  {Zilberberg}, \citenamefont {Romito},\ and\ \citenamefont
  {Gefen}}]{Zilberberg2012}%
  \BibitemOpen
  \bibfield  {author} {\bibinfo {author} {\bibfnamefont {O.}~\bibnamefont
  {Zilberberg}}, \bibinfo {author} {\bibfnamefont {A.}~\bibnamefont {Romito}},
  \ and\ \bibinfo {author} {\bibfnamefont {Y.}~\bibnamefont {Gefen}},\ }\href
  {\doibase 10.1088/0031-8949/2012/T151/014014} {\bibfield  {journal} {\bibinfo
   {journal} {Physica Crypta}\ }\textbf {\bibinfo {volume} {T151}},\ \bibinfo
  {pages} {014014} (\bibinfo {year} {2012})}\BibitemShut {NoStop}%
\bibitem [{ban()}]{bandwidth}%
  \BibitemOpen
  \href@noop {} {}\bibinfo {note} {In Sec.~\ref{derivationOfRates}, we see that for our case where $\delta
\mathcal{D}=\mathcal{D}$, one of the backaction mechanisms of the QPC
on the dot system is to give the dot level a finite width $\hbar
\mathcal{D}$. By increasing $\mathcal{D}$ one can give the detector a
sufficiently large bandwidth to detect single cotunneling events,
$\tau_{\rm{cot}} \mathcal{D} \gg 1$, where  $\tau_{\rm{cot}}
\sim \hbar/(\mu_D - \epsilon_d)$
from the Heisenberg's uncertainty principle.
This implies that, as soon as $\mathcal{D}$ is large enough to give a large
bandwidth to the detector, the induced level broadening becomes comparable with the dot's energetic distance from the leads.
Hence the QPC smears the dot level into an incoherent sequential transport regime through QPC-assisted transport.}
\BibitemShut {Stop}%
\bibitem [{\citenamefont {Hewson}(1997)}]{hewson1997kondo}%
  \BibitemOpen
  \bibfield  {author} {\bibinfo {author} {\bibfnamefont {A.~C.}\ \bibnamefont
  {Hewson}},\ }\href@noop {} {\emph {\bibinfo {title} {The Kondo Problem to
  Heavy Fermions}}},\ (\bibinfo  {publisher} {Cambridge
  University Press, Cambridge},\ \bibinfo {year} {1997})\BibitemShut {NoStop}%
\end{thebibliography}
%

\end{document}